\documentclass[preprint]{emulateapj}
\def\agt{\mathrel{\raise.3ex\hbox{$>$}\mkern-14mu\lower0.6ex\hbox{$\sim$}}}
\def\alt{\mathrel{\raise.3ex\hbox{$<$}\mkern-14mu\lower0.6ex\hbox{$\sim$}}}

\newcommand{\beq}{\begin{equation}}
\newcommand{\eeq}{\end{equation}}
\newcommand{\beqn}{\begin{eqnarray}}
\newcommand{\eeqn}{\end{eqnarray}}

\def\bL{\hbox{$\,{\cal L}\!\!\!$--}}

\def\bm#1{{\hbox{\boldmath $#1$}}}

\begin{document}
\title{
Formation of black hole and accretion disk in a massive high-entropy stellar core collapse}
\author{Yuichiro Sekiguchi \altaffilmark{1,2}
and Masaru Shibata \altaffilmark{2}}
\affil
{\altaffilmark{1} 
Department of Theoretical Astronomy, National Astronomical Observatory of
Japan, Mitaka, Tokyo 181-8588, Japan \\
\altaffilmark{2} 
Yukawa Institute for Theoretical Physics, Kyoto University, Kyoto, 
606-8502, Japan}
\begin{abstract}
We present the first numerical result of fully general relativistic axisymmetric 
simulations for the collapse of a rotating high-entropy stellar core to a black hole and
an accretion disk. The simulations are performed taking into account the
relevant microphysics. We adopt as initial condition a spherical core with constant 
electron fraction ($Y_e = 0.5$) and entropy per baryon $s$ = 8 $k_B$, and 
angular velocity is superimposed.
In the early phase, the core collapses in a homologous manner. 
Then, it experiences a weak bounce due to the gas pressure of free nucleons. 
Because the bounce is weak, the core collapses eventually to a black hole.
Subsequent evolution depends on initial angular velocity. 
When the rotation is not fast, a geometrically thin (but optically thick) accretion disk 
is formed, and shock waves are formed in the inner part of the disk.
For the moderately rotating case, the thin accretion disk expands eventually to be a
geometrically thick torus after sufficient accumulation of the thermal energy generated
at the shocks. Furthermore, convection occurs inside the torus.
Neutrino luminosities vary violently with time because of the convective motion.
For the rapidly rotating case, by contrast, a geometrically thick torus is formed
soon after the black hole formation, and convective activity is weak due to
the presence of epicyclic mode.
\end{abstract}

\section{Introduction}

Gamma-ray bursts (GRBs) have been one of the most outstanding phenomena
in the universe since their discovery in 1967 \citep{Klebesadel73}
because of their huge energy emitted in a short timescale
(isotopic equivalent luminosities of $10^{49}$--$10^{52}$ ergs/s in 
short duration of $\sim 0.01$--1000 s) and in addition, violent time 
variability of $\delta t \sim 1$ ms in time 
profiles of gamma-ray emission. 
GRBs are basically divided, in terms of their duration, into short 
bursts (SGRBs), for which duration is shorter than 2 s, and long 
bursts (LGRBs), for which duration is longer than 2 s.
Recent observations have found GRBs with overlapped features of 
the two populations \citep{Gehrels06,Gal-Yam06},
and it is also suggested that a new classification may be necessary 
\citep{Zhang09,Lv10}.
However, the large amount of energy release, short duration, and variability
timescale indicate that GRBs may be universally associated with accretion
processes onto a compact object of stellar-mass size \citep{Piran99}. 
Because a rotating black hole is the most efficient converter of gravitational binding energy 
in nature, it is now widely believed that many of central engines of GRBs are composed of 
a rotating black hole surrounded by a massive and hot accretion disk.

Although progenitors of GRBs have not been fully clarified yet,
there are accumulating observational evidences that LGRBs are
associated with collapse of massive stars \citep{WoosleyB06}.
(For reviews on progenitors of SGRBs, see e.g., \citet{Nakar07} and \citet{Lee07}).
The first solid evidence for the connection between LGRBs and 
supernovae came from spectroscopic identification of a supernova 
component (SN2003dh) in the afterglow of GRB030329 \citep{Hjorth03,Stanek03,Kawabata03}.
To date, at least six other connections
between LGRBs and supernovae have been reported:
GRB980425 with SN1998bw \citep{Galama98,Kulkarni98};
XRF020903 \citep{Soderberg05};
GRB021211 with SN2002lt \citep{DellaValle03}
GRB031203 with SN2003lw \citep{Malesani04,Cobb04,Thomsen04,Gal-Yam04};
GRB050525a with SN2005nc \citep{DellaValle06b}
and GRB060218 with SN2006aj \citep{Campana06,Pian06,Mirabal06,Modjaz06,Sollerman06};
All the GRB-associated supernovae are TypeIb/c. 
In addition, there are a wide variety of circumstance evidences \citep{WoosleyB06}:
E.g., 
observed association of afterglows of LGRBs with star forming 
regions in their host galaxies \citep{Christensen04,Fruchter06,Savaglio09,Svensson10},
and late time bumps resembled supernova components in light curves of LGRBs
\citep{Zeh04,Zeh05,Zeh06}.

The observational association between GRBs and supernovae has provided 
strong support to a scenario, so-called collapsar model, in which 
LGRBs are assumed to be originated in the collapse of a massive stellar 
core to a black hole \citep{Woosley93}.
\citet{MacFadyen99} outlined possible scenarios of driving LGRBs.
In the collapsar model, a central core of a massive star
is required to be rotating rapidly enough that 
a massive accretion disk can be formed around a black hole.
Then, pair annihilation of neutrinos emitted from 
the accretion disk to electron-positron pairs could supply sufficient 
energy to induce relativistic outflows \citep{Eichler89,Meszaros92,Narayan92,Mochkovitch93}.
The relativistic outflows are expected to form a GRB fireball.
In addition, it is suggested that strong magnetic fields of
order $10^{15}$ G, if they are present, could play an active role
in driving the relativistic outflows \citep{Nakamura92,Narayan92,Lyutikov06}.

There are three possible varieties in collapsar model \citep{Heger03}:
In Type I \citep{MacFadyen99} and Type II \citep{MacFadyen01} collapsar models, 
a proto-neutron star is assumed to be formed initially and a shock wave is launched. 
Then, in the Type I collapsar, the proto-neutron star collapses promptly to 
a black hole because the shock wave is weak, while in the Type II collapsar, 
a black hole is formed by a fallback process long after the proto-neutron star formation.
In the Type III collapsar model \citep{Heger03,Fryer01}, a black hole is
directly formed without formation of proto-neutron star.

Recently, two LGRBs (GRB060505 and GRB060614) which are not likely to be
accompanied by a supernova were discovered \citep{Fynbo06,Gehrels06,Gal-Yam06,DellaValle06a}.
The host galaxy of GRB0600505 is a star-forming galaxy similar to that of canonical LGRBs.
Such LGRBs might be associated with the Type I or Type III collapsar.
Note that there is debate about the lack of supernova feature in GRB06014 
\citep{Cobb06,Dado08} and it has been discussed that the duration of GRB060505 is
about 4 second and it may be a short GRB \citep{Ofek07}.

Because the observed supernovae associated with LGRBs are Type Ib/c and the 
relativistic jets have to reach the stellar surface \citep{Zhang04},
the progenitors should have lost their envelope before the onset of stellar core collapse; 
otherwise a peculiar evolution path is required. 
Due to these reasons, the progenitors of LGRBs are now believed to be 
rapidly rotating massive Wolf-Rayet (WR) stars.
However, ordinary WR stars are known to be accompanied by strong stellar
winds driven by radiation pressure which lead to a rapid spin-down of
the stellar core.
Here, a serious problem concerning collapsar model is that according to stellar
evolution calculations, it is very difficult to produce pre-collapse cores
which satisfy both the requirement of collapsar model and the association of 
Type Ib/c supernova, if magnetic torques and standard mass-loss rates are 
taken into account \citep{WoosleyH06}.

To resolve the above dilemma, several models have been proposed 
(see \citet{Fryer07} for a review).
\citet{Izzard04} and \citet{Podsiadlowski04} proposed binary-interaction 
models, in which the tidal force in a close binary keeps a helium star 
in synchronous, rapid rotation.
\citet{vandenHeuvel07} showed that a helium star in a close binary
with a compact companion (i.e., neutron star or black hole) can retain
sufficient angular momentum to form a progenitor of a GRB.
\citet{Fryer05} suggested a binary-merger model and showed that
a merger of two helium cores during the common-envelope inspiral phase
can produce a rapidly rotating core which satisfies the requirement of 
the collapsar models.

On the other hand, Yoon \& Langer \citep{Yoon05,Yoon06a,Yoon06b} and 
Woosley \& Heger \citep{WoosleyH06}
recently showed that a single star can fulfill the requirements of
the collapsar models if it is initially rapidly rotating 
($\gtrsim 50$\% of the Keplerian velocity at the equatorial surface) and
of low metallicity ($Z/Z_{\odot} \lesssim 0.1$).
Note that the low metallicity could keep the stellar radius smaller and
also reduce the mass loss \citep{WoosleyH06}. Both effects suppress 
the loss of angular momentum from the star.
The rapid rotation results in a short mixing timescale, which could help achieving 
a chemically homogeneous state throughout the hydrogen burning phase.
In this case, a single star could become a rapidly rotating WR star
without losing the hydrogen envelope through the stellar wind, 
avoiding the red giant phase that otherwise would cause a
significant decrease of the core angular momentum due to
magnetic torques \citep{Yoon06a}.
It is also noted that the chemically homogeneous evolution is likely to occur 
for the tidally spun-up star in a binary system \citep{Cantiello07}.

There are several supports to the chemically-homogeneous-evolution model.
Recent observations have indicated that LGRBs may prefer a low metallicity 
environment \citep{Fruchter06,Stanek06,Modjaz08,Svensson10}.
If the binary merger model resulted in most of the 
LGRB progenitors, such dependency would not be found.

Gravitational collapse of population III (Pop III) stars, 
which are assumed to be formed from metal-free gas, may be accompanied by LGRB at 
a very high redshift \citep{Schneider02,Bromm06}.
Numerical simulations have suggested that Pop III stars would be
predominantly very massive with $M \gtrsim 100 M_{\odot}$ 
\citep{Omukai01,Omukai03,Nakamura01,Abel02,Bromm02}.
Such a massive star may collapse directly to a black hole 
without producing supernova explosion (Type III collapsar).

In addition, an attempt to constrain the characteristics of LGRB progenitors
has been made by \citet{Campana08}, who 
studied in detail an absorption pattern in the X-ray spectrum of 
GRB060218 and found an extremely low O/N ratio in the surrounding 
of the progenitor, reaching a conclusion that only a progenitor star characterized by
a fast rotation and subsolar metallicity could explain it.

All of the above progenitor models of LGRBs are anomalous in the sense
that they are different from the progenitors of ordinary supernovae.
Qualitatively speaking, the progenitor models should produce a core of larger angular 
momentum than the ordinary supernova cores. 
Also, the central entropy of the core would be higher than the 
ordinary supernova cores because of its high mass:
The chemically homogeneous models tend to predict a well-mixed, larger core
with higher central entropy than the ordinary supernova core. 
It is also expected that the object formed after the binary merger will have a 
higher entropy, if the mass ratio of merging stars is not far from unity 
\citep{Suzuki07,Gaburov08}.
Thus, LGRB progenitor cores may be modeled by a rapidly rotating,
higher-entropy core, regardless of their formation processes.
Based on this assumption, in this paper, we perform collapse simulations of 
a very massive stellar core with a fairly high value of entropy 
($s=8k_{B}$ per baryon) to study effects of higher-entropy.

A number of hydrodynamic simulations have been performed for studying gravitational
collapse of such rapidly rotating, higher-entropy core in the context of 
collapsar model: 
for the Type I collapsar model, see 
\citet{Proga03}, \citet{Fujimoto06}, \citet{Dessart08}, \citet{Nagataki09}, \citet{Harikae09}, 
\citet{Lopez-Camara09}, and \citet{Ott11}; 
for the Type II collapsar model, see \citet{MacFadyen01};
for the Type III collapsar model, see \citet{Fryer01}, \citet{Shibata02},
\citet{Sekiguchi07}, \citet{Suwa07b}, and \citet{Liu07}).
Most of the simulations were performed in the Newtonian or pseudo-Newtonian gravity 
\citep{MacFadyen01,Fryer01,Proga03,Fujimoto06,Suwa07b,Dessart08,Harikae09,Lopez-Camara09}.
In such simulations, inner regions of core ($r \lesssim 5$--$20 r_{\rm S}$ where 
$r_{\rm S}$ is the Schwarzschild radius) are excised, and consequently, increase of 
the overall efficiency of accretion according to the black hole spin from 
$\approx 6$\% (zero spin) to $\approx 42$\% (maximal spin) cannot be taken into account. 
The black hole spin has significant effects on structure of the accretion disk, 
because it dramatically changes the spacetime metric near the black hole, where most of 
accretion power is released \citep{Chen07}.

Also, to guarantee formation of a centrifugally supported accretion disk 
at radii larger than the excised radius, most of Newtonian studies adopted 
angular momentum distributions that are well above the threshold of the disk formation:
The specific angular momentum $j$ for a large fraction of the core is assumed to be 
much larger than that at the innermost stable circular orbit (ISCO), $j_{\rm ISCO}$.
In such cases, gravitational energy will not be effectively
converted into thermal energy due to the large radii. Rather these models rely on
subsequent hypothetical viscous heating for generating large amount of energy.
By constant, \citet{Lee06} performed simulations of low angular momentum 
accretion flows into a black hole in the Newtonian framework. 
They found that a thin accretion disk is formed for $j\lesssim 1.9 r_{\rm S}c$
while a thick torus is formed for $j\lesssim 2.1 r_{\rm S}c$
(see also \citet{Lopez-Camara09}). \citet{Harikae09} also found similar results.

To self-consistently follow formation of a black hole and a surrounding disk,
a fully general relativistic simulation for the collapse of rapidly rotating 
massive star was first performed by \citet{Shibata02}.
Unfortunately, they could not follow the subsequent
evolution of an accretion disk around the black hole.
\citet{Sekiguchi07} and \citet{Liu07} performed 
fully general relativistic simulations of collapsar, successfully following 
formation of an accretion disk and an early evolution of the disk.
Recently, \citet{Ott11} performed simulations in the context of the collapsar 
scenario and extracting the gravitational wave signature from it.
\citet{Nagataki09} performed a long-term general relativistic
simulation in a fixed Kerr black hole background.
However, in these general relativistic simulations, 
relevant microphysical processes such as neutrino cooling were
not taken into account.

In this paper, we for the first time report the results of fully general 
relativistic simulations for the collapse of a rapidly rotating, high-entropy core, 
taking into account detailed microphysics;
a nuclear-theory-based finite-temperature equation of state (EOS), weak interaction processes
such as electron capture and pair-neutrino processes, and
neutrino cooling.
We focus on self-consistently clarifying the formation process of a rotating
black hole and surrounding accretion disk, and subsequent long-term evolution
of this system. We will show how the black hole is formed and evolved, and also
clarify the physical condition for the disk or torus 
in the vicinity of the black hole. 
In particular, this is the first work that clarifies the 
geometrical structure, thermal property (such as chemical composition, chemical
potentials, and entropy), neutrino optical depth, and neutrino luminosities of
the accretion disk in the framework of full general relativity.

The paper is organized as follows. We first briefly summarize
the basic equations, the input physics, and numerical setup in
Section \ref{Sec_Setting}. The main results are described in
Section \ref{Sec_Results}. Discussion of our results together with 
prospects for GRB production are given in Section \ref{Sec_Discussion}. 
Section \ref{Sec_Summary} is devoted to a summary.
Throughout this paper, $\hbar$, $k_{B}$, $c$, and $G$ denote 
the Planck's constant, the Boltzmann's constant, the velocity of light, and
the gravitational constant, respectively.
We adopt the geometrical unit $c=G=1$ in Sections \ref{Sec_3+1} and 
\ref{Sec_Leakage}, which is commonly used in numerical relativity.

\section{Setting}\label{Sec_Setting}

\subsection{Einstein's equation and gauge conditions} \label{Sec_3+1}

The standard variables in the 3+1 decomposition of Einstein's equation are 
the three-dimensional metric $\gamma_{ij}$ and 
the extrinsic curvature $K_{ij}$
on the three-dimensional hypersurface defined by \citep{York79}
\beqn
\gamma_{\mu\nu} &\equiv& g_{\mu\nu} + n_{\mu}n_{\nu}, \\
K_{\mu\nu} &\equiv& - \frac{1}{2} \bL _{n} \gamma_{\mu\nu}, 
\eeqn
where $g_{\mu\nu}$ is the spacetime metric,
 $n_{\mu}$ is the unit normal to a three-dimensional
hypersurface, and $\bL _{n}$ is the Lie derivative with
respect to the unit normal $n^{\mu}$. 
Then we can write the line element in the form
\beq
ds^{2} = - \alpha^{2} dt^{2} + \gamma_{ij}(dx^{i}+\beta^{i}dt)
(dx^{j}+\beta^{j}dt),
\eeq
where $\alpha$ and $\beta^{i}$ are the lapse function
and the shift vector which describe the gauge degree of freedom.

Numerical simulation is performed in the BSSN formulation \citep{Shibata95,Baumgarte99}
in which the spatial metric $\gamma_{ij}$ is conformally decomposed as
$\gamma _{ij} = e^{ 4\phi}\tilde{\gamma}_{ij}$
where the condition $\det (\tilde{\gamma}_{ij}) = 1$ is imposed for the
conformal metric $\tilde{\gamma}_{ij}$. 
From this condition, the conformal factor is written as
$\phi = \frac{1}{12}\ln \gamma$ and $ \gamma \equiv \det(\gamma_{ij})$. 
The extrinsic curvature $K_{ij}$ is decomposed into the trace part 
$K$ and the traceless part $A_{ij}$ as
$K_{ij} = A_{ij} + (1/3)\gamma_{ij}K$ . 
The traceless part $A_{ij}$ is conformally
decomposed as $A_{ij} = e^{4\phi}\tilde{A}_{ij}$.
Consequently, the fundamental quantities for the evolution equation are now
split into 
$\phi, \tilde{\gamma}_{ij}$, $K$, and $\tilde{A}_{ij}$.
Furthermore, the auxiliary variable 
$ F_{i} \equiv \delta^{jk}\partial_{k}\tilde{\gamma}_{ij} $ 
is introduced in the BSSN formulation \citep{Shibata95}.

To stably follow the spacetime after appearance of a black hole,
we evolve $W \equiv e^{-2\phi}$ instead of $\phi$ following 
\citet{Marronetti08}. The primary reason is that $\phi$ diverges
at the center of a black hole in the vertex-center grid.
With the choice of $W$, such pathology can be avoided, as first pointed
out by \citet{Campanelli06}, in which $\chi \equiv e^{-4\phi}$ was
used instead of $W$. Merits of using $W$  are that (i) the equation 
for the Ricci tensor is slightly simplified, 
(ii) no singular term appears in the evolution equations even for 
$W \rightarrow 0$, and (iii) the determinant of $\gamma_{ij}$ is
always positive \citep{Marronetti08,Yamamoto08}.

We assume axial and equatorial symmetries of the spacetime and 
the so-called Cartoon method \citep{Shibata00,Shibata03a,Alcubierre01b} 
is adopted to avoid possible problems around the coordinate singularities of 
the cylindrical coordinates.
In the present code, we use a 4th-order finite difference scheme in 
the spatial direction and a 3rd-order Runge-Kutta scheme in the time integration.
The advection terms such as $\beta^{i}\partial_{i}\phi$ are evaluated
by a 4th-order upwind scheme \citep{Brugmann08}.

As the gauge conditions for the lapse, 
we use a dynamical slicing \citep[cf.][]{Alcubierre01a}:
\beq
\partial_{t} \alpha = -2K \alpha.
\eeq
It is known that this dynamical slicing enables to perform a long-term
evolution of neutron stars as well as has a strong singularity avoidance
property in the black hole spacetime.
The shift vector is determined by solving the following dynamical 
equation \citep{Shibata03b}
\begin{equation}
\partial_{t}\beta^{k} = \tilde{\gamma}^{kl} (F_{l} + \Delta t
 \partial_{t} F_{l}).
\label{Dynbeta}
\end{equation}
Here the second term in the right-hand side is necessary for
numerical stability, and $\Delta t$ denotes the numerical timestep.

\subsection{Hydrodynamic equations coupled to general relativistic 
leakage scheme}\label{Sec_Leakage}

Recently, \citet{Sekiguchi10a,Sekiguchi10b} developed a fully general relativistic 
hydrodynamic code implementing a nuclear-theory-based finite-temperature EOS,
self-consistent electron and positron captures, and neutrino cooling by
a general relativistic leakage scheme. Neutrino heating is not included in 
the current version of leakage scheme. 
Since we assume the axial and equatorial symmetry of the spacetime,  
the hydrodynamics equations are solved in the cylindrical coordinates 
$(\varpi, \varphi,z)$ where $\varpi = \sqrt{x^{2}+y^{2}}$.
We follow \citet{Sekiguchi10b} for a solution of the hydrodynamic equations to which 
the readers may refer for the details.
In this section, we adopt the geometrical unit $c=G=1$.

\subsubsection{Energy-momentum conservation equation}

The basic equations of general relativistic hydrodynamics with neutrinos
are
\beq
\nabla_{\alpha}(T^{\rm Total})^{\alpha}_{\ \beta} 
= \nabla_{\alpha}\left[(T^{\rm F})^{\alpha}_{\ \beta} +
(T^{\nu})^{\alpha}_{\ \beta} \right] = 0, \label{Eq_Ttot}
\eeq
where $(T^{\rm Total})_{\alpha \beta}$ is the total energy-momentum
tensor, and $(T^{\rm F})_{\alpha \beta}$ and $(T^{\nu})_{\alpha \beta}$
are the energy-momentum tensor of fluids and neutrinos, respectively.
Following \citet{Sekiguchi10b}, the neutrino energy-momentum tensor is decomposed
into 'trapped-neutrino' ($(T^{\nu,{\rm T}})_{\alpha\beta}$) 
and 'streaming-neutrino' ($(T^{\nu,{\rm S}})_{\alpha\beta}$) parts as
\beq
(T^{\nu})_{\alpha\beta} = (T^{\nu,{\rm T}})_{\alpha\beta} +
                        (T^{\nu,{\rm S}})_{\alpha\beta}. 
\label{Eq_Tnu_div}
\eeq
Here, the trapped-neutrino part phenomenologically represents neutrinos which 
interact sufficiently frequently with matter, and the streaming-neutrino 
part describes a phenomenological flow of neutrinos streaming out of the system.
\citet{Lieb09} developed a more sophisticate method in terms 
of the distribution functions of trapped and streaming neutrinos 
in the Newtonian framework.

Streaming-neutrinos are produced with a leakage rate 
$Q^{\rm leak}_{\alpha}$, according to
\beq
\nabla_{\alpha}(T^{\nu,{\rm S}})^{\alpha}_{\ \beta} =  Q^{\rm leak}_{\beta}.
\label{Eq_T_nuS}
\eeq
On the other hand, the trapped-neutrino part is combined with the 
fluid part as
\beq
T_{\alpha\beta} \equiv (T^{\rm F})_{\alpha\beta} 
+ (T^{\nu,{\rm T}})_{\alpha\beta}. \label{Eq_T_comb}
\eeq
Then the equation for $T_{\alpha \beta}$ is
\beq
\nabla_{\alpha}T^{\alpha}_{\ \beta} = -Q^{\rm leak}_{\beta} \label{Eq_T}.
\eeq
We solve Eqs. (\ref{Eq_T_nuS}) and (\ref{Eq_T}) for the 
energy-momentum conservation equation.

The energy-momentum tensor of the fluid and trapped-neutrino parts
($T_{\alpha \beta}$) is treated as that of the perfect fluid,
\beq
T_{\alpha\beta} = (\rho + \rho \varepsilon + P)
 u_{\alpha}u_{\beta} + P g_{\alpha\beta}, \label{T_fluid}
\eeq
where $\rho$ and $u^{\alpha}$ are the rest mass density and the 4-velocity.
The specific internal energy density ($\varepsilon$) and the pressure ($P$) 
are the sum of contributions from the baryons 
(free protons, free neutrons, $\alpha$-particles, and heavy nuclei), 
leptons (electrons, positrons, and {\it trapped-neutrinos}), and photons as,
\beqn
P &=& P_{B} + P_{e} + P_{\nu} + P_{ph}, \\
\varepsilon &=&  
\varepsilon_{B} + \varepsilon_{e} +
\varepsilon_{\nu} + \varepsilon_{ph} ,
\eeqn
where subscripts '$B$', '$e$', '$ph$', and '$\nu$' denote the components
of baryons, electrons and positrons, photons, and trapped-neutrinos, 
respectively. 

The streaming-neutrino part, on the other hand, is set to be a general
form of
\beq
(T^{\nu,{\rm S}})_{\alpha\beta}= 
E n_{\alpha}n_{\beta} + F_{\alpha}n_{\beta} + F_{\beta}n_{\alpha} + P_{\alpha\beta},
\label{T_neutrino}
\eeq
where $F_{\alpha}n^{\alpha}=P_{\alpha \beta}n^{\alpha}=0$. 
In order to close the system,
we need an explicit expression of $P_{\alpha \beta}$.
In this paper, we adopt a simple form 
$P_{\alpha \beta}=\chi E \gamma_{\alpha \beta}$ with $\chi = 1/3$.
Then we solve Eq. (\ref{Eq_T_nuS}) in a high resolution shock capturing 
scheme \citep{Sekiguchi10b}.

The closure relation employed in this paper is not very physical.
Also, recall that we do not consider the so-called neutrino heating 
in this paper. To treat the neutrino heating accurately, 
a more sophisticated closure relation is required. However, such a 
study is beyond the scope of this 
paper. A more sophisticated treatment of neutrino transport 
equations, together with incorporating the neutrino heating,
will be needed in the future \citep[e.g.,][]{SKSS11}.

\subsubsection{Lepton-number conservation equations}

The conservation equations of the lepton fractions are written 
schematically as
\beqn
&&\!\! \frac{d Y_{e}}{dt} = -\gamma_{e} , \label{dYe} \\
&&\!\! \frac{d Y_{\nu_{e}}}{dt} = \gamma_{\nu_{e}},  \label{dYnu} \\ 
&&\!\! \frac{d Y_{\bar{\nu}_{e}}}{dt} = \gamma_{\bar{\nu}_{e}},  \label{dYna} \\ 
&&\!\! \frac{d Y_{\nu_{x}}}{dt} = \gamma_{\nu_{x}},  \label{dYno}  
\eeqn
where $Y_{e}$, $Y_{\nu_{e}}$, $Y_{\bar{\nu}_{e}}$, and $Y_{\nu_{x}}$ denote
the fractions per baryon number for electrons, electron neutrinos, electron
anti-neutrinos, and $\mu$ and $\tau$ neutrinos and anti-neutrinos,
respectively. Here we consider, as local reactions, the electron
capture, the positron capture, electron-positron pair annihilation, plasmon decay, 
and the Bremsstrahlung radiation of
pair neutrinos, where $\nu$ and $\bar{\nu}$ denote the three flavors of neutrinos and
anti-neutrinos.

The source terms are given by 
\beqn
&&\gamma_{e} = \gamma_{\nu_{e}}^{\rm local} - \gamma_{\bar{\nu}_{e}}^{\rm local}, \\
&&\!\! \gamma_{\nu_{e}} = \gamma_{\nu_{e}}^{\rm local} - \gamma_{\nu_{e}}^{\rm leak}, \\
&&\!\! \gamma_{\bar{\nu}_{e}} = \gamma_{\bar{\nu}_{e}}^{\rm local} 
                        - \gamma_{\bar{\nu}_{e}}^{\rm leak}, \\
&&\!\! \gamma_{\nu_{x}} = \gamma_{\nu_{x}}^{\rm local} - \gamma_{\nu_{x}}^{\rm leak}, 
\eeqn
where $\gamma^{\rm local}$'s and $\gamma^{\rm leak}$'s are
the local production and leakage rates of each species of neutrinos, respectively.
Because $\gamma^{\rm local}_{\nu}$\ 's are characterized 
by the timescale of weak-interaction processes 
$t_{\rm wp}\sim \vert Y_{e}/\dot{Y}_{e} \vert$ which can be much shorter
than the dynamical timescale \citep[e.g.,][]{Bruenn85}, a straightforward explicit 
solution of Eqs. (\ref{dYe})--(\ref{dYno}) leads, in general, to a numerical 
instability.
Therefore we follow the procedure proposed in \citet{Sekiguchi10b} 
to solve the equations stably in an explicit manner. 

First, in each timestep $n$, the conservation equation of 
the {\it total} lepton fraction ($Y_{l}=Y_{e}-Y_{\nu_{e}}+Y_{\bar{\nu}_{e}}$), 
\beqn
&&\!\! \frac{d Y_{l}}{dt} = -\gamma_{l},  \label{dYl} 
\eeqn
is solved together with the conservation equation of $Y_{\nu_{x}}$, Eq. (\ref{dYno}),
in advance of solving the whole of the lepton conservation 
equations (Eqs. (\ref{dYe}) -- (\ref{dYno})).
Then, assuming that the $\beta$-equilibrium is achieved, 
values of the lepton fractions in the $\beta$-equilibrium ($Y_{e}^{\beta}$,
$Y_{\nu_{e}}^{\beta}$, and $Y_{\bar{\nu}_{e}}^{\beta}$) are calculated from
the evolved value of $Y_{l}$. 

Second, regarding $Y_{\nu_{e}}^{\beta}$ and $Y_{\bar{\nu}_{e}}^{\beta}$ as the
maximum allowed values of the neutrino fractions in the next 
timestep $n+1$,  the source terms are limited so that each value of $Y_{\nu}$'s in 
the timestep $n+1$ cannot exceed that of $Y_{\nu}^{\beta}$ 's.
This limiter procedure enables to solve explicitly the whole of the lepton 
conservation equations (Eqs. (\ref{dYe}) -- (\ref{dYno})).

Third, the following conditions are checked,
\beqn
\mu_{p}+\mu_{e} < \mu_{n}+\mu_{\nu_{e}} , \\
\mu_{n}-\mu_{e} < \mu_{p}+\mu_{\bar{\nu}_{e}},
\eeqn
where $\mu_{p}$, $\mu_{n}$, $\mu_{e}$, $\mu_{\nu_{e}}$, and $\mu_{\bar{\nu}
e}$ are the chemical potentials of protons, neutrons, electrons, 
electron neutrinos, and electron anti-neutrinos, respectively. 
If both conditions are satisfied, the values of
the lepton fractions in the timestep $n+1$ are set to be those in 
the $\beta$-equilibrium value;  
$Y_{e}^{\beta}$, $Y_{\nu_{e}}^{\beta}$, and $Y_{\bar{\nu}_{e}}^{\beta}$.
On the other hand, if either or both conditions are not satisfied, 
the lepton fractions in the timestep $n+1$ is set to be those obtained
by solving the whole of the lepton-number conservation equations.

\subsection{Microphysics} \label{Sec_MicroPhys}

\subsubsection{Equation of state}

In this paper, we employ a tabulated EOS derived by \citet{Shen98}, 
which is based on the Br\"uckner-Hartree-Fock-type relativistic mean field theory.
The maximum gravitational mass of a cold spherical neutron star in this EOS 
is much larger than the canonical neutron star mass $\approx 1.4M_{\odot}$
as $\approx 2.2 M_{\odot}$ \citep{Shen98}.
The framework of the relativistic mean field theory 
is extended with the Thomas-Fermi spherical cell model approximation 
to describe not only the homogeneous matter but also 
an inhomogeneous one.

The thermodynamical quantities of dense matter
at various sets of $(\rho, Y_{p}, T)$ are
calculated to construct the numerical data table for simulation.
Here $Y_{p}$ is the total proton fraction per baryon number. 
The original table covers a range of density
$10^{5.1}$--$10^{15.4}$ g/cm$^{3}$, proton fraction $0.0$--$0.56$,
and temperature $0$--$100$ MeV, which are required for supernova
simulation. The original table has been extended to higher 
density \citep{Sumiyoshi07,Sumiyoshi08} 
and higher temperature \citep{Nakazato08} ranges of $10^{5.1}$--$10^{17}$ g/cm$^{3}$ 
and $0$--$400$ MeV, which are required for following black hole formation
\citep{Sumiyoshi06}.

It should be noted that the causality is guaranteed to be satisfied in
this framework, whereas the sound velocity
sometimes exceeds the velocity of the light in the non-relativistic
framework, e.g., in the EOS by \citet{Lattimer91}. 
This is one of the benefits of the relativistic EOS.

To consistently calculate the pressure and the internal energy of 
electrons and positrons, the charge neutrality condition $Y_{p} = Y_{e}$ should 
be solved to determine the electron chemical potential
$\mu_{e}$ for each value of the baryon rest mass density $\rho$ and
the temperature $T$ in the EOS table.
Namely, it is required to solve the equation
\beq 
n_{e}(\mu_{e},T) \equiv n_{-} - n_{+} = \frac{\rho Y_{e}}{m_{u}}
\label{n_to_mu}
\eeq
in terms of $\mu_{e}$ for given values of $\rho$, 
$T$, and $Y_{e}\ (= Y_{p})$.
Here, $m_{u} = 931.49432$ MeV is the atomic mass unit, 
and $n_{-}$ and $n_{+}$ are the total number densities
(i.e., including electron-positron pairs) of
electrons and positrons, respectively. 
Then, assuming that electrons and positrons obey the Fermi-Dirac distribution, 
the number density, the pressure, and the internal
energy density of electrons and positrons are calculated in a standard
manner \citep[e.g.,][]{Cox68}.

The pressure and the specific internal energy density of photons are
given by
\beqn
P_{r} = \frac{a_{r}T^{4}}{3},\ \ 
\varepsilon_{r} = \frac{a_{r}T^{4}}{\rho},
\eeqn
where $a_{r}=(\pi^{2}k_{B}^{4})/(15c^{3}\hbar^{3})$ is the radiation constant.

In this paper, trapped-neutrinos are assumed to interact sufficiently frequently 
with matter that be thermalized. Therefore they are described as ideal Fermi gases 
with the matter temperature. From the numerically evolved neutrino 
fractions $Y_{\nu}^{\rm evol}$, the chemical potentials of neutrinos ($\mu_{\nu}$) are 
calculated by solving
\beq
Y_{\nu}^{\rm evol} = Y_{\nu}(\mu_{\nu}, T) 
= \frac{m_{u}}{\rho}n_{\nu}(\mu_{\nu}, T).
\eeq
Then the pressure and the internal energy of trapped-neutrinos are 
calculated in the same manner as for electrons, 
using $\mu_{\nu}$ and the matter temperature.

\subsubsection{Weak interaction and leakage rate}

Following \citet{Sekiguchi10b}, the leakage rates are defined by
\beqn
&&\!\! Q_{\nu}^{\rm leak}= (1-e^{-b\tau_{\nu}}) Q_{\nu}^{\rm diff} 
+ e^{-b\tau_{\nu}} Q_{\nu}^{\rm local}, \label{Q_leak} \\
&&\!\! \gamma_{\nu}^{\rm leak}= (1-e^{-b\tau_{\nu}}) \gamma_{\nu}^{\rm diff} 
+ e^{-b\tau_{\nu}} \gamma_{\nu}^{\rm local}, \label{g_leak}
\eeqn
where $\tau_{\nu}$ is the optical depth of neutrinos and $b$ is a parameter
which is typically set as $b^{-1}=2/3$. The optical depth can be
computed from the cross sections following an often employed prescription
\citep{Ruffert96,Rosswog03}:
The optical depth is calculated by 
\beq
\tau_{\nu} = {\rm min}\left[ \tau^{\varpi}_{\nu}, \tau^{z}_{\nu}, \tau^{r}_{\nu} \right],
\eeq
where $\tau_{\nu}^{\varpi}$, $\tau_{\nu}^{z}$, and $\tau_{\nu}^{r}$ are the optical 
depths along $\varpi$, $z$, and the radial directions, respectively. 
We calculate, for example, $\tau_{\nu}^{z}$ by
\beq
\tau_{\nu}^{z}(\varpi,z) = \int_{z}^{z_{\rm out}}\kappa_{\nu}(\varpi, z')dz',
\eeq 
where $\kappa_{\nu}$ is the opacity and $z_{\rm out}$ denotes the outer 
boundary in the $z$-direction.
$\tau_{\nu}^{\varpi}$ and $\tau_{\nu}^{r}$ are calculated in a similar manner.

Then, because $Q^{\rm leak}_{\nu}$ should be regarded as the emissivity of
neutrinos measured in the {\it fluid rest frame}, $Q^{\rm
leak}_{\alpha}$ is defined as \citep{Shibata_etal07,Sekiguchi10a,Sekiguchi10b}
\beq
Q^{\rm leak}_{\alpha} = Q^{\rm leak}_{\nu}u_{\alpha}.
\eeq\label{leakage_source_Q}

As the local production reactions of neutrinos, we consider
the electron and positron captures ($\gamma_{\nu_{e}}^{\rm ec}$ and
$\gamma_{\bar{\nu}_{e}}^{\rm pc}$) following \citet{Fuller85},
the electron-positron pair annihilation
($\gamma_{\nu_{e} \bar{\nu}_{e}}^{\rm pair}$ for electron-type neutrinos
and $\gamma_{\nu_{x} \bar{\nu}_{x}}^{\rm pair}$ for the other type)
following \cite{Cooperstein86},
the plasmon decays
($\gamma_{\nu_{e} \bar{\nu}_{e}}^{\rm plas}$ and
$\gamma_{\nu_{x} \bar{\nu}_{x}}^{\rm plas}$) following \citet{Ruffert96},
and the Bremsstrahlung processes
($\gamma_{\nu_{e} \bar{\nu}_{e}}^{\rm Brems}$ and
$\gamma_{\nu_{x} \bar{\nu}_{x}}^{\rm Brems}$) following \citet{Burrows06}.
Then, the local reaction rates for the neutrino fractions are
\beqn
&& 
\gamma_{\nu_{e}}^{\rm local} = \gamma_{\nu_{e}}^{\rm ec} + 
\gamma_{\nu_{e} \bar{\nu}_{e}}^{\rm pair} + 
\gamma_{\nu_{e} \bar{\nu}_{e}}^{\rm plas} +
\gamma_{\nu_{e} \bar{\nu}_{e}}^{\rm Brems}, \label{gnlocal}\\
&& 
\gamma_{\bar{\nu}_{e}}^{\rm local} = \gamma_{\bar{\nu}_{e}}^{\rm pc} + 
\gamma_{\nu_{e} \bar{\nu}_{e}}^{\rm pair} + 
\gamma_{\nu_{e} \bar{\nu}_{e}}^{\rm plas} +
\gamma_{\nu_{e} \bar{\nu}_{e}}^{\rm Brems}, \label{galocal}\\
&& 
\gamma_{\nu_{x}}^{\rm local} = 
\gamma_{\nu_{x} \bar{\nu}_{x}}^{\rm pair} + 
\gamma_{\nu_{x} \bar{\nu}_{x}}^{\rm plas} +
\gamma_{\nu_{x} \bar{\nu}_{x}}^{\rm Brems}. \label{gxlocal}
\eeqn
Similarly, the local neutrino energy emission rate $Q_{\nu}^{\rm local}$ is given
by
\beqn
Q_{\nu}^{\rm local} = Q_{\nu_{e}}^{\rm ec} + Q_{\bar{\nu}_{e}}^{\rm pc} 
&+& 2\,(Q_{\nu_{e} \bar{\nu}_{e}}^{\rm pair} + 
        Q_{\nu_{e} \bar{\nu}_{e}}^{\rm plas} +
        Q_{\nu_{e} \bar{\nu}_{e}}^{\rm Brems}) \nonumber \\
&+& 4\,(Q_{\nu_{x} \bar{\nu}_{x}}^{\rm pair} + 
        Q_{\nu_{x} \bar{\nu}_{x}}^{\rm plas} +
        Q_{\nu_{x} \bar{\nu}_{x}}^{\rm Brems})\ . \label{Qlocal}
\eeqn
The explicit forms of the local rates in
Eqs. (\ref{gnlocal})--(\ref{Qlocal}) are found in \citet{Sekiguchi10b}.

We follow the recent work by \citet{Rosswog03} for
the diffusive neutrino emission rates $\gamma_{\nu}^{\rm diff}$ and 
$Q_{\nu}^{\rm diff}$ in Eqs. (\ref{Q_leak}) and (\ref{g_leak}).
The explicit forms of $\gamma_{\nu}^{\rm diff}$ and $Q_{\nu}^{\rm diff}$
are found in \citet{Sekiguchi10b}.

\subsection{Initial model} \label{Sec_Init}

\begin{figure}
  \begin{center}
    \epsscale{1.0}
    \plotone{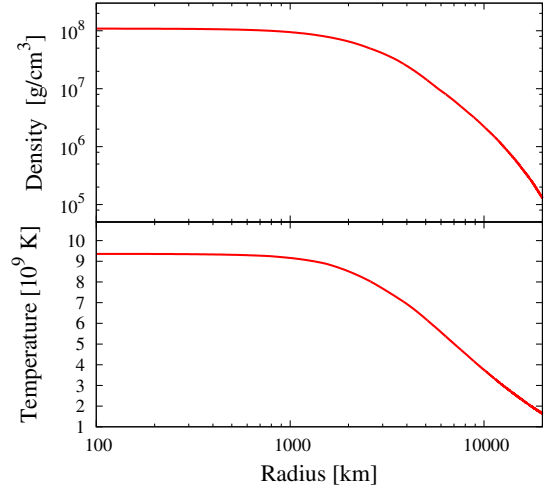}
  \end{center}
\caption{Radial profiles of density (upper panel) and temperature (lower panel)
of the initial configuration.\label{fig_init_prof}}
\end{figure}

\begin{figure}
  \begin{center}
    \epsscale{1.0}
    \plotone{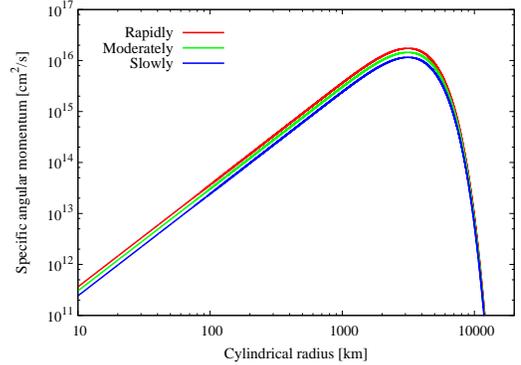}
  \end{center}
\caption{Radial profiles of specific angular momentum for 
the slowly, moderately, and rapidly rotating models.
\label{fig_init_spej}}
\end{figure}

\begin{figure}
  \begin{center}
    \epsscale{1.0}
    \plotone{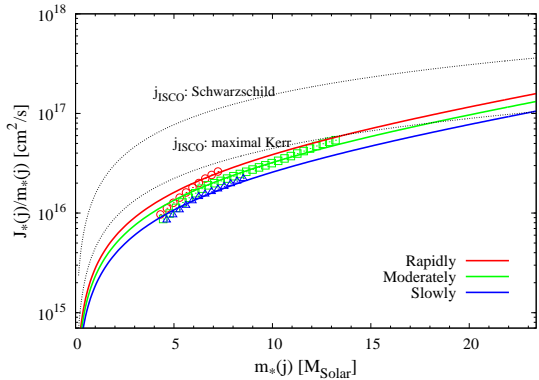}
  \end{center}
\caption{Distributions of the averaged specific angular momentum 
for slowly (blue curve), moderately (green curve), and 
rapidly (red curve) rotating models.
The specific angular momentum required to support a fluid element in a circular orbit 
at ISCO around a Schwarzschild black hole and a maximally rotating Kerr black hole 
of mass $m_{*}(j)$ is shown together (black dotted curves).
The blue triangles, green squares, and red circles indicate the numerical results for the 
path along which specific angular momentum and mass of the black hole 
formed in the collapse of the slowly, moderately, and rapidly rotating models
follow, respectively (see Sec. \ref{Sec_Results}).
\label{fig_spej}}
\end{figure}

\begin{figure}
  \begin{center}
    \epsscale{1.0}
    \plotone{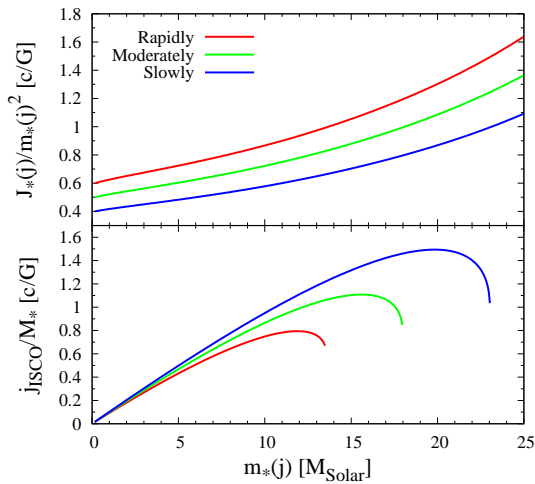}
  \end{center}
\caption{The spin parameter distribution, $q_{*}(j)$, and the specific angular momentum 
at ISCO, $j_{\rm ISCO}(j)/M_{*}$, in units of $c/G$, 
where $M_{*}$ is the total baryon mass.
\label{fig_jISCO}}
\end{figure}

Because there are no realistic models of rotating progenitors 
derived by multi-dimensional pre-collapse evolution calculations or no 
binary progenitor models, 
we prepare approximate initial models in the following manner \citep{Nakazato07}.
We first calculate a spherical equilibrium configuration with a constant
electron fraction of $Y_{e}=0.5$ and with a constant entropy per baryon
$s=8 k_{B}$. 
We set the central density to be $\rho_{c} \approx 10^{8}$ g/cm$^{3}$.
The corresponding central temperature is $T_{c}\approx 9\times 10^{9}$ K,
which is higher than the critical temperature for the 
photo-dissociation of heavy nuclei to occur.
Following \citet{Nakazato07}, we define the outer boundary of the 'iron 
core' to be where the temperature is $5 \times 10^{9}$ K. 
Note that most of heavy nuclei in inner parts of this 'iron core' in fact are
already photo-dissociated.
Then the mass and the radius of 
the core are $M_{\rm iron} \approx 13 M_{\odot}$ and $r_{\rm iron} \approx 7000$ km.
In numerical simulation we follow a region of $r_{\rm tot} \approx 14000$ km 
($> r_{\rm iron}$) in which the total mass of $M_{\rm tot} \approx 23 M_{\odot}$ is enclosed.
The radial profiles of density and temperature are shown in Figure \ref{fig_init_prof}.

For the purpose of reference, we note that our initial model might correspond 
to entropy per baryon for a star with initial mass of 
$\approx$ 120--$130M_{\odot}$ \citep{Bond84}.
However, a recent study \citep{Waldman08} predicts that such massive stars will undergo a 
pulsational pair instability and considerable mass loss, resulting in hydrostatic
degenerate iron cores of mass with $\sim 3 M_{\odot}$, which is different from the
initial model adopted in this paper. 
The 300 $M_{\odot}$ progenitor used by \citet{Fryer01} has a central entropy of 
$\sim 8 k_{B}$ per baryon. However such a very massive model does not form an
iron core in hydrostatic fashion, but rather goes unstable much earlier 
burning phase. 
Note that these are results in a spherical single star with solar metallicity. 
Anomalous stars, such as stars in interacting binary and Pop III stars, might form 
such high-entropy cores \citep{Nakazato07}. 

Little is also known about the angular momentum distribution
in the progenitor core. Thus, we employ the following rotation profile
\beq
\Omega(\varpi) = \Omega_{0} 
\exp \left[-\frac{1}{2}\frac{R_{c}^{2}}{(\varpi^{2} + R_{c}^{2})} \right] 
\exp \left[-\frac{\varpi^{2}}{R_{0}^{2}}\right] ,
\eeq
where $\varpi = \sqrt{x^{2} + y^{2}}$, and
$\Omega$, $R_{0}$, and $R_{c}$ are parameters which control 
the degree of differential rotation. The exponential cut-off factor is
introduced by a practical reason for numerical simulation:
if the specific angular momentum in the outer region of the core is too large,
the matter escapes from the computational domain.
However, the most part of the 'iron core' is almost uniformly rotating.
We fix the values of $R_{0}$ and $R_{c}$ as 
$R_{0}=r_{\rm tot}/5$ and $R_{c}=r_{\rm tot}/8$, respectively.
We vary $\Omega_{0}$ as 0, 0.4, 0.5 and 0.6 rad/s (hereafter referred to as
spherical, slowly rotating, moderately rotating, and rapidly rotating models).
The rotation period in the central region is $\approx 10$--15 s.
This is by one order of magnitude longer than the dynamical timescale 
$(G\rho_{c})^{-1/2} \sim 0.4$ s. Thus, the progenitor star is not assumed
to be rapidly rotating. 
The profiles of specific angular momentum
along the cylindrical radius are plotted in Figure \ref{fig_init_spej}.

Figure \ref{fig_spej} plots an averaged specific angular momentum distribution
defined by $J_{*}(j)/m_{*}(j)$. 
Here, $j$ is the specific angular momentum of a fluid element, which is a
conserved quantity in axially symmetric spacetime in the absence of viscosity.
$m_{\ast}(j)$ is a rest mass distribution as a function of $j$,
which is the integrated baryon rest mass of fluid elements with the
specific angular momentum less than $j$, defined by \citep{Shibata02}
\beq
m_{\ast}(j) \equiv 2\pi \int _{j'< j} \rho_{\ast} r^{2}dr d(\cos \theta).
\eeq
Similarly, $J_{*}(j)$ is an angular momentum distribution defined by
\beq
J_{\ast}(j) \equiv 2\pi \int _{j'< j} \rho_{\ast} j' r^{2}dr d(\cos \theta).
\eeq
These conserved quantities are often used in general 
relativistic study to predict a possible outcome of the collapse
\citep{Shibata02,Shapiro04,Sekiguchi04}.

It should be noted that the specific angular momentum considered in this paper 
is rather small for a large fraction of fluid elements, in the sense that it is 
smaller than the angular momentum required for a fluid element to stay outside the 
innermost stable circular orbit (ISCO), $j_{\rm ISCO}$, around a Schwarzschild 
black hole. In this sense, our model is 'sub-Keplerian'.
This is by contrast with many of previous models in which 
the specific angular momentum of well above $j_{\rm ISCO}$ is usually
imposed (e.g., \citet{MacFadyen99}, but see \citet{Lee06},
\citet{Lopez-Camara09}, and \citet{Harikae09}).
In the present condition, the fluid elements of such small specific angular
momentum form a black hole, while those of large specific angular momentum
does a disk (torus).
 
Now, to infer the evolution of a black hole surrounded by accreting materials,
let us consider ISCO around a hypothetical black hole located at the center.
If the value of $j$ of a fluid element is smaller than that at the ISCO, 
$j_{\rm ISCO}$, for the hypothetically formed black hole, the fluid element 
will fall into the seed black hole eventually. 
The value of  $j_{\rm ISCO}$ will change as the ambient fluid elements 
accrete into the black hole.
If $j_{\rm ISCO}$ increases as a result of the accretion, 
more ambient fluid elements will fall into the black hole.
On the other hand, if $j_{\rm ISCO}$ decreases during the accretion,
the accretion into the black hole will be suppressed, and then, the black hole
will approach to a quasi-stationary state with a small accretion rate.

To estimate the value of $j_{\rm ISCO}$, we assume that the spacetime
metric can be instantaneously approximated by that of a Kerr spacetime
of mass $m_{*}(j)$ and the non-dimensional spin parameter 
$q_{*}(j)\equiv c J_{*}(j)/G m_{*}(j)^{2}$.
On these approximations, we may compute $j_{\rm ISCO}$ of a 
black hole \citep[e.g.,][]{Shapiro83}.

For all the models considered in this paper, $q_{*}(j)$ is smaller
than unity for a fraction of fluid elements with small specific angular momentum.
As a result of this fact, these fluid elements can form a black hole in the dynamical 
timescale. However, this will not be the case for the initial condition with 
$q_{*}(j) > 1$ in an inner region.
In this case, a black hole will not be formed directly because the Kerr space time
with the spin parameter greater than unity contains a naked singularity.
Instead, a rotating oblate object will be the outcome \citep{Saijo09,Sekiguchi04}.
Such an oblate object will be unstable against nonaxisymmetric deformation, and then,
angular momentum will be transported by the hydrodynamic torque from the
inner region to the outer one. As a result of a sufficient amount of 
angular momentum transport, a black hole will be eventually formed \citep{Zink07}.
This suggests that the timescale for black hole formation may be determined
by the timescale for the angular momentum transport.
We do not consider this possibility in this paper.

Figure \ref{fig_jISCO} plots the spin parameter distribution ($q_{*}(j)$)
and $j_{\rm ISCO}(j) = j_{\rm ISCO}[ m_{*}(j), q_{*}(j) ]$ as functions of $m_{*}(j)$.
This figure clearly indicates that the value of $j_{\rm ISCO}(j)$ takes the maximum
at $m_{*}(j) \approx 12, 16,$ and 20$M_{\odot}$ for the rapidly,
moderately, and slowly rotating models, respectively.
These values show a possible final value of black hole mass, which is smaller
than the total mass of the system.
This indicates that a certain fraction of the material with mass $> M_{\odot}$ will
form a disk around the black hole.
It should be noted that the curves of Figures \ref{fig_spej} and \ref{fig_jISCO}
indicate the possible evolution path of the black hole only approximately.
In determining $j_{\rm ISCO}$ as a function of $m_{*}(j)$, we assume that
a fluid element of smaller value of $j$ falls into black hole earlier.
However, this is not always the case in the dynamical evolution of the system,
because the material in the outer region near the rotation axis has a small
value of $j$ and falls into the black hole in a late time.

\subsection{Analysis of black hole and accretion disk} \label{Sec_BH}

The formation of a black hole is ascertained by finding apparent horizon 
\citep{Shibata97}.
Then, we calculate two geometrical quantities which possibly characterize mass 
of a black hole. One is an irreducible mass defined by
\beq
M_{\rm irr} = \frac{c^{2}}{G}\sqrt{\frac{A_{H}}{16\pi}},
\eeq
where $A_{H}$ is the area of the apparent horizon. 
The other mass is associated with the circumference proper length 
along the equatorial surface $C_{e}$:
\beq
M_{ce} = \frac{c^{2}}{G}\frac{C_{e}}{4\pi}.
\eeq
This should agree with the mass of a Kerr black hole in the stationary
axisymmetric spacetime.
Note that in the case of a Schwarzschild black hole $M_{\rm irr}=M_{ce}$.

We also estimate black hole mass using an approximate conservation law,
\beq
M_{\rm con} = M_{\rm ADM} - M_{*,r>r_{AH}},
\eeq
where $M_{\rm ADM}$ is the ADM mass of the system and $M_{*,r>r_{\rm AH}}$ is the
rest mass of baryons located outside the apparent horizon.
It is suggested that $M_{ce}$ may be a good indicator of mass of a black hole 
even in the presence of a massive accretion disk \citep{Shibata07}.
As we shall see in Section \ref{Sec_Results}, $M_{ce}$ and $M_{\rm con}$
agree approximately with each other, and thus, we use $M_{ce}$ as the black hole mass, namely,
\beq
M_{\rm BH} \equiv M_{ce} \approx M_{\rm con}.
\eeq

The non-dimensional spin parameter $q$ of a Kerr 
black hole can be calculated from the ratio between polar and equatorial 
circumferential radii of event horizon, $C_{p}$ and $C_{e}$, 
\beq 
\frac{C_{p}}{C_{e}} = \frac{\sqrt{2\hat{r}_{+}}}{\pi}
\int_{0}^{\pi/2}d\theta \sqrt{1-\frac{q^{2}}{2\hat{r}_{+}}\sin^{2}\theta},
\label{Eq_cp-ce}
\eeq
where $\hat{r}_{+}=1+\sqrt{1-q^{2}}$.
The definition of $M_{\rm irr}$ for a Kerr black hole,
\beq
\frac{M_{\rm irr}}{M_{\rm BH}} = \sqrt{\frac{1}{2}
\left(
1 + \sqrt{1-q^{2}}
\right)}
\label{Eq_Chris}
\eeq
may be also used to estimate the black hole spin.
However, by contrast with $M_{ce}$, $C_{p}/C_{e}$ and $M_{\rm irr}/M_{\rm BH}$ 
are not very good indicators of the black hole spin when a massive disk 
presents \citep{Shibata07}. In the case of equilibrium configuration of
a black hole surrounded by a massive disk, it was found that a spin 
parameter estimated by Eqs. (\ref{Eq_cp-ce}) and (\ref{Eq_Chris}) 
decreases with the increase of 
disk mass and with the decrease of the inner edge of a disk.
Accordingly, a spin parameter estimated by Eqs. (\ref{Eq_cp-ce}) and (\ref{Eq_Chris}) 
may contain an error of $\Delta q \sim 0.1$ because a massive accretion disk 
falling into a black hole is formed in the present study.

We note that we approximately calculate $C_{p}$, $C_{e}$, $M_{\rm BH}$, and $M_{\rm irr}$
measuring the geometrical quantities of apparent horizon.
The disagreement between the event horizon and the apparent 
horizon may be large if the spacetime is not stationary, e.g., during the mass
accretion phase in which the black hole mass dynamically increases.
This fact makes the reliability of these methods worse.
It should be noted that the dynamical horizon formalism \citep[e.g.,][]{Schnetter06} 
could be used to obtain more reliable estimation for mass and angular momentum of 
a dynamical black hole.

Instead of using Eqs. (\ref{Eq_cp-ce}) and (\ref{Eq_Chris}), we estimate angular momentum of 
a black hole using the conservation law,
\beq
J_{\rm BH} \equiv J_{\rm con} = J_{\rm tot} - J_{r>r_{AH}} - \Delta J_{\nu},
\eeq
where $J_{\rm tot}$ is the total angular momentum of the system,
$J_{r>r_{\rm AH}}$ is the amount of angular momentum located outside 
the apparent horizon, and $\Delta J_{\nu}$ is the amount of 
angular momentum carried away by neutrinos. 
We here ignore a small contribution of $\Delta J_{\nu}$.
Then, we adopt the quantity
\beq
q_{\rm BH} \equiv \frac{c J_{\rm con}}{G M_{\rm BH}^{2}}
\eeq
as an approximate indicator of the non-dimensional 
spin parameter of a black hole.

An accretion disk will be formed in the collapse of the rotating models.
Because it is difficult to strictly define disk mass, we approximately 
estimate it by 
\beq
M_{\rm disk} \equiv \int_{\rho > \rho_{\rm cut},\, r_{\rm AH}<r<r_{\rm cut}}\rho_{*}d^{3}x,
\eeq
where $\rho_{\rm cut}$ is a cutoff density which characterizes density 
near the surface of the accretion disk, $r_{\rm AH}$ is radius of 
apparent horizon, and $r_{\rm cut}$ is a cutoff radius which characterize 
the size of the accretion disk. 
Although $M_{\rm disk}$ is no more than an approximate indicator,
the disk mass may be estimated by $M_{\rm disk}$ in a reasonable accuracy:
When $\rho_{\rm cut}$ is larger than the surface density, slight change
of $\rho_{\rm cut}$ will result in large change of $M_{\rm disk}$.
By contrast, in the case that $\rho_{\rm cut}$ is smaller than the 
surface density, $M_{\rm disk}$ will not change much even if $\rho_{\rm cut}$ 
is decreased to some extent, because density outside the disk is low.
We choose $\rho_{\rm cut}$ so that $M_{\rm disk}$ is not largely affected by
a small change in $\rho_{\rm cut}$ and typically set $\rho_{\rm cut} = 10^{10}$ g/cm$^{3}$. 

In this paper, we basically consider two rates, mass accretion rate into a black hole 
($\dot{M}_{\rm BH}$) and mass infalling rate onto an accretion disk 
($\dot{M}_{\rm disk}$), which are associated with time evolution of $M_{\rm BH}$ 
and $M_{\rm disk}$, respectively. The total mass infalling rate onto the
system of a black hole surrounded by an accretion rate is then approximately 
given by $\dot{M} = \dot{M}_{\rm BH} + \dot{M}_{\rm disk}$.

\subsection{Grid Setting}\label{Sec_Grid}

\begin{table*}
 \begin{center}
  \caption{Summary of the regridding procedure}\label{regrid} 
  \begin{tabular}{c|cccccc} \hline
     & $\Phi_{c} \le 0.0125  $ & $  \le \Phi_{c} \le 0.025 $ & $  \le \Phi_{c} \le 0.05 $ 
     & $\le \Phi_{c} \le 0.1 $ & $\Phi_{c} \le 0.2         $ & $\Phi_{c} \ge 0.2 $ \\
     \hline
 $\Delta x_{0}$  (km) & 10.1  & 4.8    & 2.2   & 0.98   & 0.45  & 0.22   \\
 $\delta$      & 0.008 & 0.0075 & 0.007 & 0.0065 & 0.006 & 0.0065  \\
 $N$           & 316   & 412    & 524   & 652    & 796   & 960     \\
 $L$ (km)      & 14600 & 13300  & 11800 & 10100  & 8700  & 7700    \\ \hline
 $\Delta x_{0}$  (km) & 5.8  & 2.5    & 1.1   & 0.47   & 0.22  & 0.097   \\
 $\delta$      & 0.0075 & 0.007 & 0.0065 & 0.006 & 0.0055 & 0.005  \\
 $N$           & 400   & 520    & 656   & 812    & 980   & 1200     \\
 $L$ (km)      & 14600 & 13300  & 11800 & 10100  & 8700  & 7700    \\ \hline
  \end{tabular}
 \end{center}
\end{table*}

In numerical simulations, we adopt a nonuniform grid, in which 
the grid spacing is increased according to the rule
\beq
d x_{j+1} = (1 + \delta) d x_{j}, \ \ \ \ d z_{l+1} = (1 + \delta) d z_{l},
\eeq
where $d x_{j} \equiv x_{j+1} - x_{j}$, $d z_{l} \equiv z_{l+1} - z_{l}$, 
and $\delta$ is a constant.
In addition, a regridding technique \citep{Shibata02,Sekiguchi05} 
is adopted to assign a sufficiently large number of grid points inside the
collapsing core, saving the CPU time efficiently.
The regridding is carried out whenever the characteristic radius
of the collapsing star decreases by a factor of 2--3. 
At each regridding, the minimum grid spacing is decreased by a factor
of $\sim 2$ and the geometrical factor $\delta$ is changed slightly.

All the quantities on the new grid are calculated
using the fifth-order Lagrange interpolation. 
However, for the fluid quantities such as $\rho$ and $h$,
the fifth-order interpolation could fail because the
interpolation may give negative values of $\rho$ and $h-1$.
In such cases, we adopt the linear interpolation
to calculate the quantities on the new grid, based on
the prescription proposed by \citet{Yamamoto08}.
In each regridding, we solve the Hamiltonian constraint
equation numerically.

To avoid discarding a large amount of the matter in the outer region 
(i.e., for approximately keeping the location of outer boundary), 
we also increase the grid number at each regridding.
For the regridding, we define 
a relativistic gravitational potential
$\Phi_c \equiv 1 -\alpha_c~ (\Phi_c>0)$ where $\alpha_c$ is the central value
of the lapse function.
Because $\Phi_c$ is approximately proportional to $M/R$ where $M$ and $R$
are characteristic mass and radius of the core, 
$\Phi_c^{-1}$ can be used as a measure of the characteristic
length scale of the stellar core for the regridding. 

To check the convergence of results, a simulation in a finer grid resolution 
is also performed.
Table \ref{regrid} summarizes the regridding parameters 
($N$ and $L$ are mesh number and computational domain) of each
level of the regridding procedure for normal (upper) and higher (lower) 
resolutions.

\section{Results}\label{Sec_Results}

\subsection{Spherical model}

\begin{figure}
  \begin{center}
    \epsscale{1.0}
    \plotone{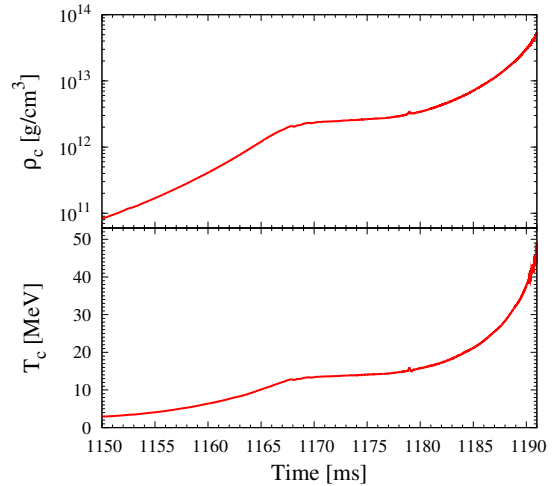}
  \end{center}
  \caption{Time evolution of the central values of density and temperature for the spherical model. 
    The collapsing core experiences weak bounce at $t\approx 1168$ ms.
    We note that apparent horizon is formed at $t\approx 1193$ ms.
    \label{fig_rho-tem}}
\vspace{3mm}
\end{figure}

\begin{figure}
  \begin{center}
    \epsscale{1.0}
    \plotone{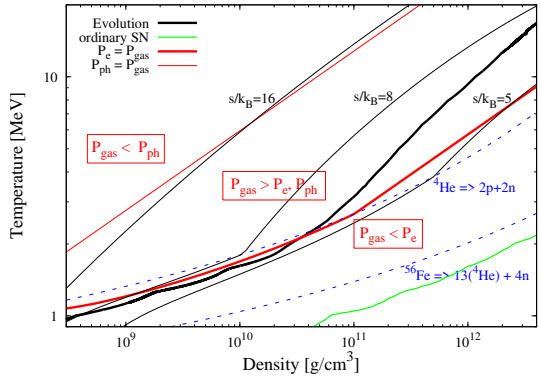}
  \end{center}
  \caption{Evolution path of the central values of density and temperature for the spherical model 
           in $\rho$-$T$ plane (thick black solid curve).
           The thick and thin red solid curves show the boundary at which the condition 
           $P_{e}=P_{\rm gas}$ or $P_{r}=P_{\rm gas}$ is satisfied.
           The thin black solid curves show evolution paths with constant entropy per baryon for
           $s/k_{B}=5$, 8 and 16. The two blue dashed curves denote the values of $(\rho, T)$ with which
           $^{56}$Fe or $^{4}$He will be half by mass due to the photo-dissociation.
           An evolution path of the central values of density and temperature for an ordinary 
           supernova core (\citealt{Sekiguchi10b}, see the text for details) is shown together 
           (solid green curve).
  \label{fig_rho-T}}
\end{figure}

\begin{figure*}
  \begin{center}
    \epsscale{1.0}
    \plotone{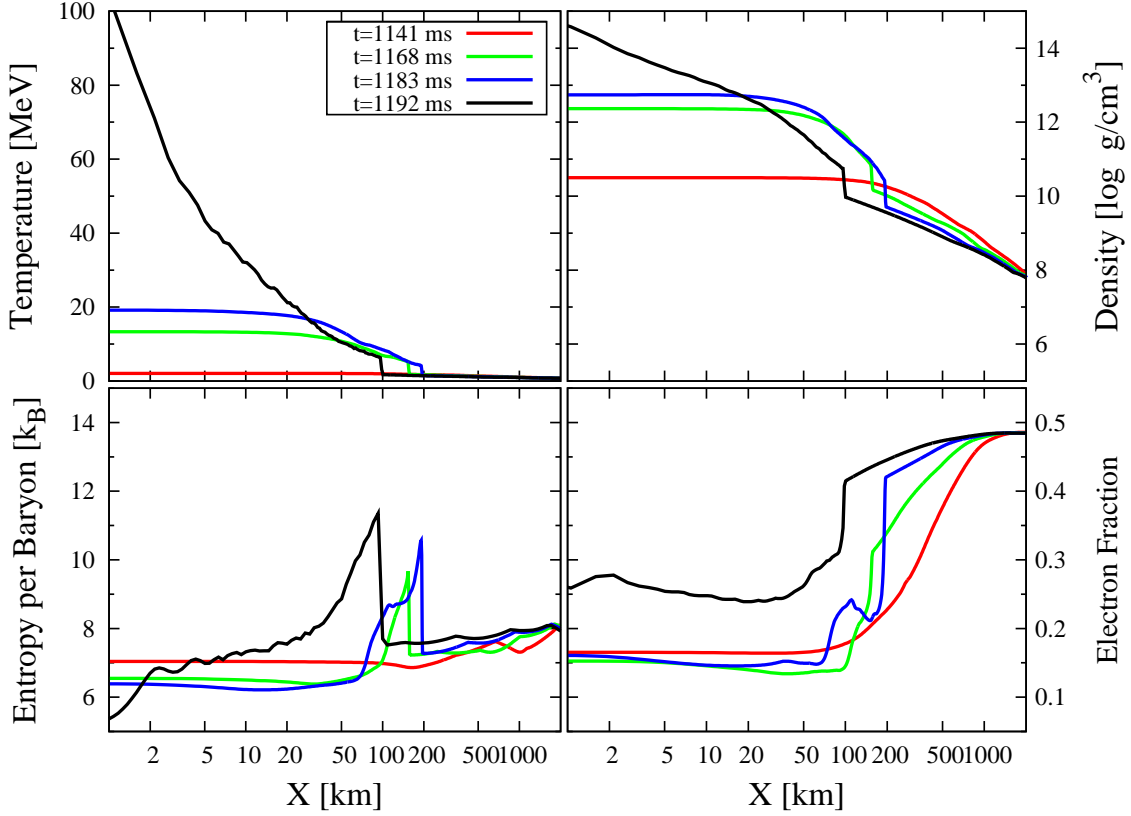}
  \end{center}
  \caption{Radial profiles of temperature, density, entropy per baryon, and electron fraction 
   along the radial coordinate in the equator at $t\approx 1141$, 1168 (bounce), 
   1179 (shock stall), and 1192 (just before the
   apparent horizon formation) ms. Formation of a shock for $t\gtrsim 1168$ ms is due to
   the weak bounce.
  \label{fig_rprofile}}
\end{figure*}

\begin{figure}
  \begin{center}
    \epsscale{1.0}
    \plotone{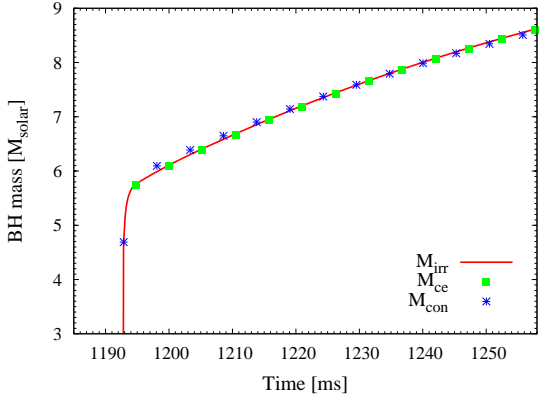}
  \end{center}
  \caption{Time evolution of black hole mass for the spherical model. \label{fig_BH-sp}}
\end{figure}

\begin{figure}
  \begin{center}
    \epsscale{1.0}
    \plotone{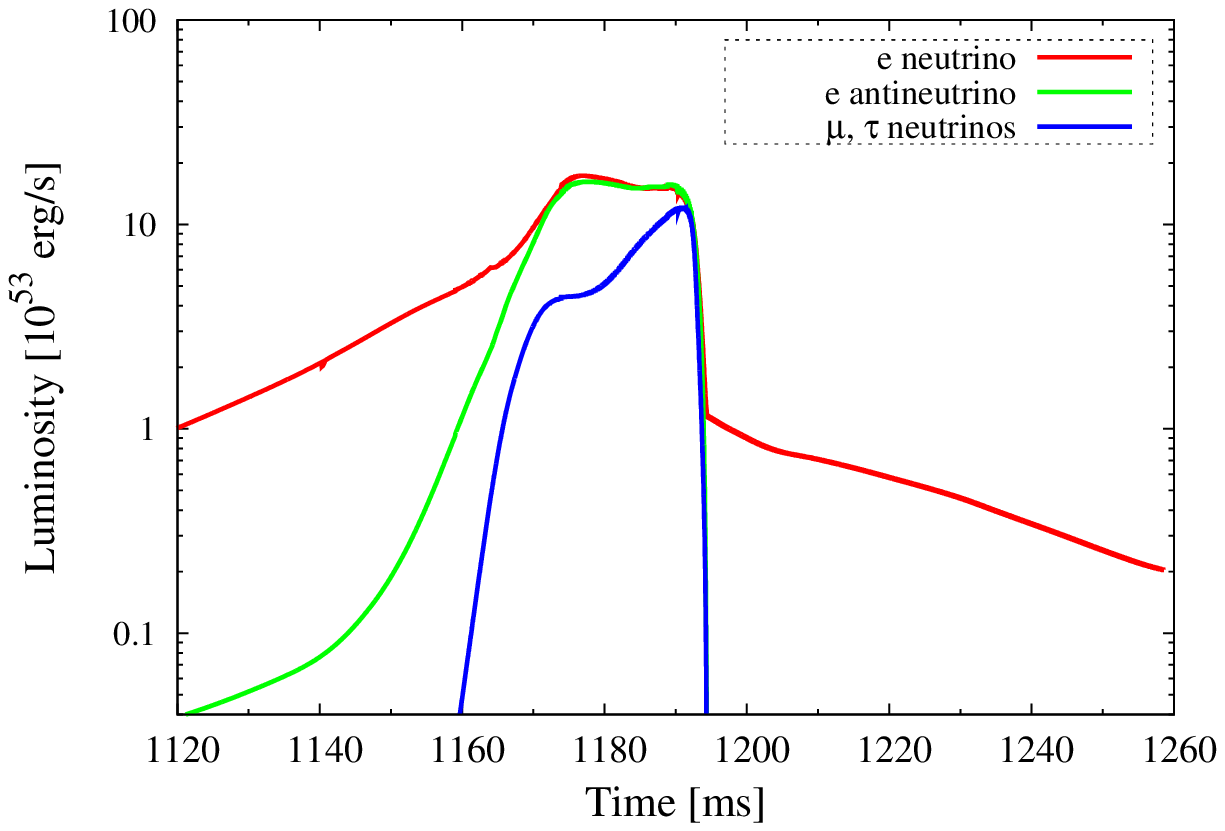}
  \end{center}
  \caption{Time evolution of neutrino luminosities for the spherical model.
    Note that the black hole is formed at $t\approx 1193$ ms.
\label{fig_lum-sp}}
\end{figure}

In this section, we describe the features of collapse dynamics for the  spherical model
as a baseline for the rotational models described later.
As in the core collapse of an ordinary supernova for which the central 
value of entropy per baryon is $s/k_{B} \sim 1$,
gravitational collapse is triggered by the electron capture and 
the photo-dissociation of heavy nuclei. 
Then the collapse in the early phase proceeds in a homologous manner.
Because of the higher value of the entropy per baryon ($s/k_{B} = 8$),
most of heavy nuclei are resolved into heliums by the photo-dissociation
(cf. Figure \ref{fig_rho-T}).
As the collapse proceeds and as a result, temperature increases, heliums
are resolved into free nucleons ($p$, $n$).
As we shall see below, due to the higher entropy and the resulting difference
in the baryon composition, the collapse dynamics in a late phase 
is different from that of an 
ordinary supernova core.

\subsubsection{Gas pressure dominated bounce}\label{Sec_Bounce}

It is known that an ordinary supernova core experiences
a bounce when the central density exceeds the nuclear density 
($\rho_{\rm nuc} \sim 2 \times 10^{14}$ g/cm$^{3}$) above which the pressure 
increases drastically due to the repulsive nuclear force.
In the present case, the collapse is not decelerated 
by the nuclear force but by the thermal gas pressure $P_{\rm gas}$ at 
a density far below $\rho_{\rm nuc}$.
Such a feature of dynamics was already reported in the recent 
simulations \citep{Fryer01,Nakazato07,Suwa07b}.
We reconfirm this previous discovery and clarify the origin of
this phenomena in more detail in the following.

The evolution of the central values of density and temperature 
for the spherical model is shown in Figure \ref{fig_rho-tem}. 
At $t\approx 1168$ ms the core experiences a weak bounce (see also Figure \ref{fig_rprofile}).
The central density at the bounce is below the nuclear density 
($\approx 2\times 10^{12}$ g/cm$^{3}$) and the central value of the temperature 
is $\approx 13$ MeV. At these values of central density and temperature, the pressure 
in the inner core is dominated by the thermal pressure of gas composed primarily of
free nucleons and heliums.

This situation is different from that for $t \lesssim 1160$ ms, for which the pressure 
in most region of the inner part is dominated by
the degenerate pressure of relativistic electrons.
Because the adiabatic index of non-relativistic gas is $\Gamma=5/3$,
which is much larger than that for relativistic degenerate electrons, $\Gamma \approx 4/3$,
the collapse is decelerated due to a sudden increase of the pressure.
The radial profiles of temperature, density, entropy, and entropy per baryon 
at the bounce along the equator are shown in Figure \ref{fig_rprofile}.
This figure shows that the profiles do not vary significantly after the bounce,
for $1168 \lesssim t \lesssim 1183$ ms.

The critical value of entropy per baryon for the gas-pressure-dominated 
bounce to occur may be approximately estimated as follows.
We plot paths along which entropy per baryon is constant in Figure 
\ref{fig_rho-T} (see the thin black curves).
For $s/k_{B} \lesssim 5$, paths of $P_{e}=P_{\rm gas}$ and constant entropy
do not intersect. For $s/k_{B} \gtrsim 16$, on the other hand,
the gas-pressure-dominated bounce cannot occur because the pressure is 
always dominated by the radiation pressure of photons 
(see the thin red curve in Figure \ref{fig_rho-T}).
Therefore, most of the results obtained in this paper would be applied 
qualitatively to models with $5\lesssim s/k_{B} \lesssim 16$.

\subsubsection{Shock stall and Black hole formation} \label{Sec_BHform}

As in the case of ordinary core collapse, a shock wave is formed at the
gas-pressure-dominated bounce, and then, it propagates outward (see Figure \ref{fig_rprofile}). 
Because this bounce is weak, the shock wave is stalled soon after
the bounce, at $t\approx 1179$ ms (cf. Figure \ref{fig_rho-tem}). 
Near the stalled shock, a region of negative gradient of electron
fraction ($\partial Y_{e}/ \partial r < 0$) is formed 
(see the blue curve in Figure \ref{fig_rprofile}) because
neutrinos carry away the lepton number from the shock-heated region.
It is known that such a configuration is unstable to convection.
However, because the thermally supported hot inner core quickly ($\sim 10$ ms) 
collapses to a black hole, convection does not play an important role by 
contrast with the case of ordinary supernovae.

Figure \ref{fig_BH-sp} plots the time evolution of black hole mass for the spherical model.
Note that the three masses of the black hole (see Section \ref{Sec_BH}) approximately 
agree with each other (see Figure \ref{fig_BH-sp}).
Apparent horizon is formed at $t\approx 1193$ ms. After the apparent horizon 
formation, we continue the simulation using a hydrodynamic excision technique 
\citep{Hawke05}, similar to adopted in \citet{Sekiguchi07}.

Black hole mass at the moment of its formation is $\approx 5.8M_{\odot}$, which is 
much larger than the maximum mass of cold spherical neutron stars 
($M_{\rm cold NS, max}\approx 2.2M_{\odot}$ for Shen's EOS).
This is because the maximum mass of a hot neutron star can be much larger
than the canonical value $M_{\rm cold NS, max}$ due to the higher entropy.
It is found that approximate average value of the entropy is $s/k_{B}\sim 7$ 
just before the black hole formation (see Figure \ref{fig_rprofile}). 
\citet{Nakazato07} calculated the maximum mass of a hot 
neutron star using Shen's EOS. 
According to their result, the maximum mass is $\approx 5.6M_{\odot}$ for 
an isentropic core of $s/k_{B}\approx 7$ with $Y_{e}=0.1$, which agrees 
approximately with our present result.
After the formation of the black hole, its mass increases gradually 
as the accretion of the material from the outer region proceeds.
In the first $\sim 100$ ms, the mass accretion rate into the black hole 
is $\dot{M}_{\rm BH} \sim 30M_{\odot}$ s$^{-1}$.

\subsubsection{Neutrino luminosities}

Figure \ref{fig_lum-sp} 
plots the time evolution of neutrino luminosities
for the spherical model. Before the weak bounce, average energy of $\mu$ and $\tau$ neutrinos
is largest and electron neutrinos are 
dominantly emitted and emissivity of electron anti-neutrinos is much smaller
because electrons are mildly degenerate with the electron degeneracy parameter 
of $\eta_{e} \sim 4 (>1)$ and the positron fraction, responsible for anti-neutrino emission, is small. 
Note that the temperature is relatively low as $T\sim$ a few MeV.
At leading order, ignoring the blocking terms due to weak degeneracy of neutrinos, 
energy emission rates associated with the electron capture and with the positron 
capture are, respectively, written as
\beqn
Q^{\rm ec}_{\nu_{e}} &\propto& X_{p}F_{5}(\eta_{e}), \\
Q^{\rm pc}_{\bar{\nu}_{e}} &\propto& X_{n}F_{5}(-\eta_{e}).
\eeqn
Here, the Fermi-Dirac integrals are approximately given by \citep[e.g.,][]{Fuller85}
\beqn
F_{5}(-\eta_{e}) &\approx& 120 e^{-\eta_{e}}, \\
F_{5}( \eta_{e}) &\approx& \frac{\eta_{e}^{6}}{6} + \frac{5\pi^{2}}{6}\eta_{e}^{4}
+ \frac{7\pi^{2}}{6}\eta_{e}^{2} + \frac{31\pi^{2}}{126} - 120 e^{-\eta_{e}},
\eeqn
which give, for $\eta_{e} \sim 4$, $F_{5}(\eta_{e}) \sim 3000$ and 
$F_{5}(-\eta_{e})\sim 2$.
For this stage, it is found $X_{p}/X_{n} \sim 0.1$ where $X_{n}$ and $X_{p}$ are the
neutron and proton fractions. 
Therefore, the relation of $Q^{\rm ec}_{\nu_{e}} \gg Q^{\rm pc}_{\bar{\nu}_{e}}$ holds.

After the weak bounce, the degeneracy parameter becomes low as 
$\eta_{e} \sim 1.5$ because high temperature of $T \gtrsim 10$ MeV is achieved.
In this case, $F_{5}(\eta_{e}) \sim 300$ and $F_{5}(-\eta_{e})\sim 30$,
and electron neutrinos and electron anti-neutrinos are approximately
identically emitted for $X_{p}/X_{n} \sim 0.1$ because 
$Q^{\rm ec}_{\nu_{e}} \sim Q^{\rm pc}_{\bar{\nu}_{e}}$.

The peak luminosities of electron neutrinos ($\approx 1.8 \times 10^{54}$ erg/s) 
and anti-neutrinos ($1.6 \times 10^{54}$ erg/s) are achieved soon after the 
bounce (at $t \approx 1176$ ms) because neutrinos in the hot postshock region,
where the density is not so large that optical depth for neutrinos is small,
are copiously emitted. 
These luminosities remain approximately constant until black hole is formed.
This happens due to the following competing effects;
as a result of neutrino emission, thermal energy in the neutrino emission 
region is decreased, whereas as a result of compression associated with
the collapse, temperature in the neutrino emission region is increased.

The peak luminosities of $\mu$ and $\tau$ neutrinos,
on the other hand, are achieved just before the black hole formation.
This is because the temperature significantly increases (see Figure \ref{fig_rprofile}) 
due to the adiabatic compression to enhance the pair production channel 
of neutrinos.
Note that pair processes of neutrino production depend strongly on 
the temperature as $Q^{\rm pair}_{\nu \bar{\nu}} \propto T^{9}$.
Just before the black hole formation, luminosities of all the
species of neutrinos become approximately identical.
This shows that the pair production process is dominant.

Soon after the black hole formation at $t\approx 1193$ ms, 
neutrino luminosities decrease drastically because the main 
neutrino-emission region is swallowed by the black hole.
For the spherically symmetric case, i.e., in the absence of an accretion 
disk formation, neutrino luminosities damp monotonically 
as the density of infalling material decreases.
The total energies emitted b neutrinos over the entire time of the
simulation are $E_{\nu,{\rm tot}} \approx 8.3 \times 10^{52}$, $5.2 \times 10^{52}$,
and $4.5 \times 10^{52}$ erg for electron neutrinos,
electron anti-neutrinos, and {\it total} of $\mu$ and $\tau$ neutrinos,
respectively.

Before closing this subsection, we briefly compare our results for the spherical model with 
those in \citet{Nakazato07}, who performed spherically symmetric general 
relativistic simulations in which the Boltzmann equation is solved for neutrino transfer 
with relevant weak interaction processes. Note that the evolution after the 
black hole formation was not followed in their simulations 
because they adopted the so-called Misner-Sharp coordinates 
\citep{Misner64}, by which evolution of black hole cannot be followed.
According to their results for a model with the initial entropy of 
$s/k_{B} = 7.5$, the maximum neutrino luminosities achieved are 
$L_{\nu_{e}} \approx L_{\bar{\nu}_{e}} \approx 8 \times 10^{53}$ erg/s and 
$L_{\nu_{x}} \approx 4 \times 10^{53}$ erg/s,
which are by a factor of 2--3 smaller than those in our results.
The primary reason for this is that their computation was finished before 
the peak luminosity is reached due to the choice of their time coordinate, 
which is not suitable for following black hole 
evolution. However, the qualitative feature of luminosity curves for 
each species of neutrinos in our simulation agrees with that in \citet{Nakazato07} for the
phase before the black hole formation.

\subsection{Moderately rotating model}\label{Sec_RotModel}

\begin{figure*}
  \begin{center}
    \epsscale{1.0}
    \plotone{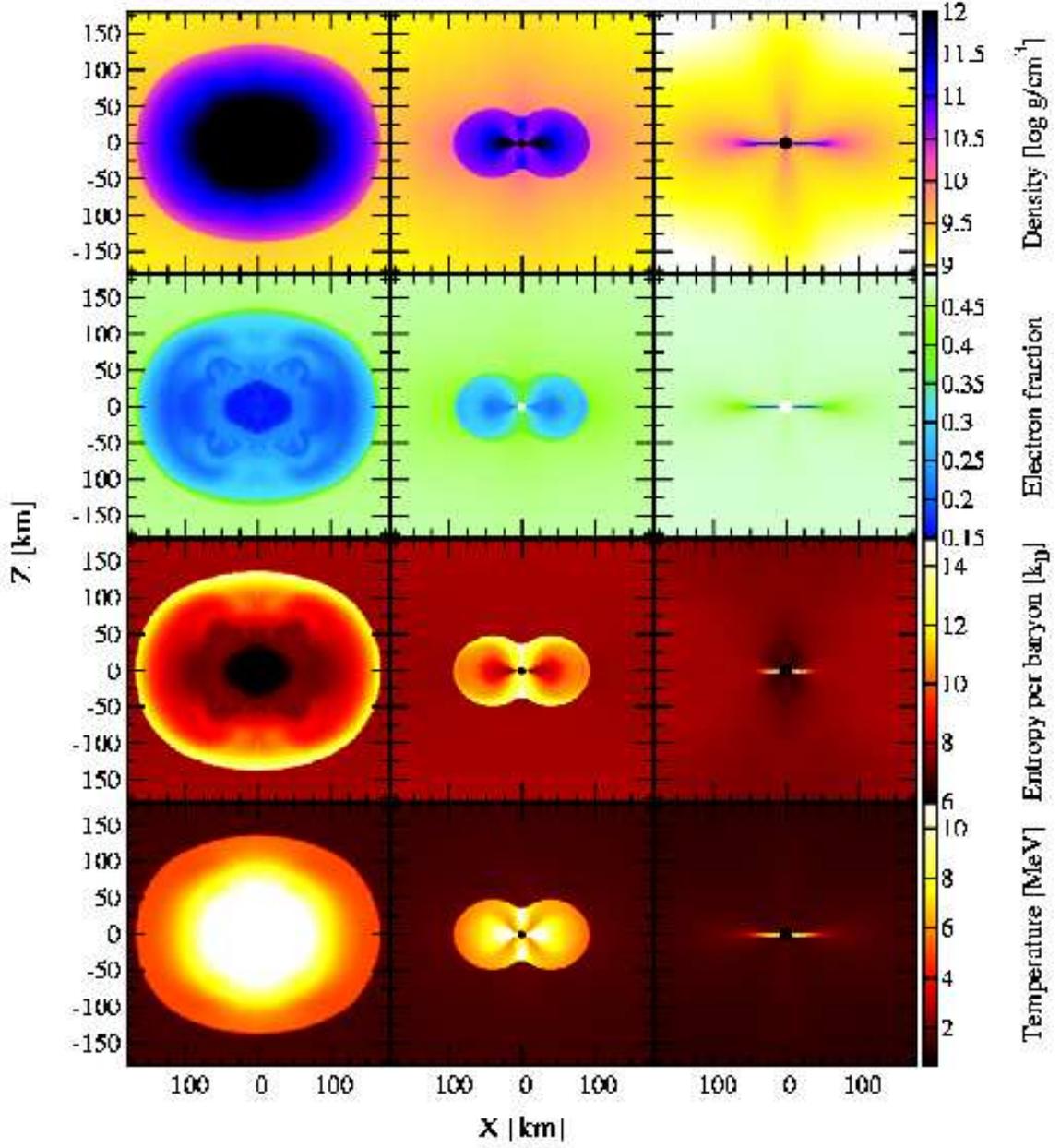}
  \end{center}
  \caption{Contours of rest mass density (panels in the first row), 
    electron fraction (panels in the second row), 
    entropy per baryon (panels in the third row), and
    temperature (panels in the fourth row) 
    at $t\approx1367$ (left panels), 1374 (middle panels), 
    and 1444 (right panels) for the moderately rotating model.
    The black regions in the contours of rest mass density
    and entropy per baryon, and the white regions in the contours of electron fraction
    at $t=1374$ and 1444 ms are inside the apparent horizon.
  \label{fig_con-disk}}
\end{figure*}

\begin{figure}
  \begin{center}
    \epsscale{1.0}
    \plotone{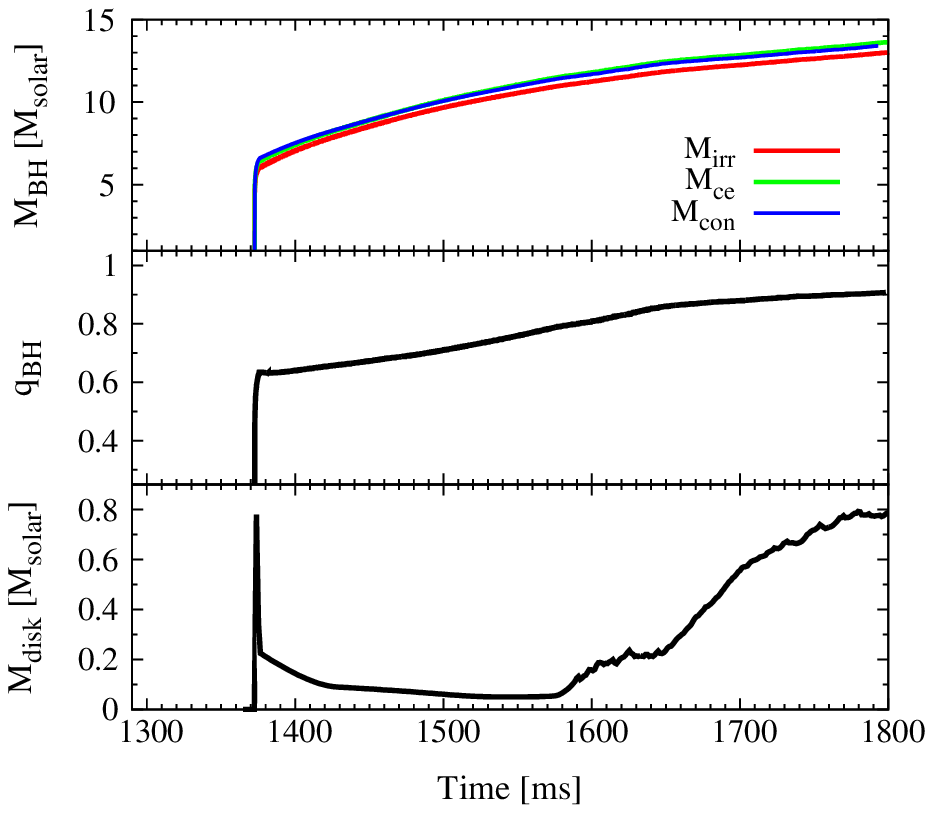}
  \end{center}
  \caption{Time evolution of mass (the top panel) and the non-dimensional 
    spin parameter (the lower panel) of the black hole and disk mass (the bottom panel) 
    for the moderately rotating model.
  \label{fig_BH-r05}}
\end{figure}

\begin{figure}
  \begin{center}
    \epsscale{1.0}
    \plotone{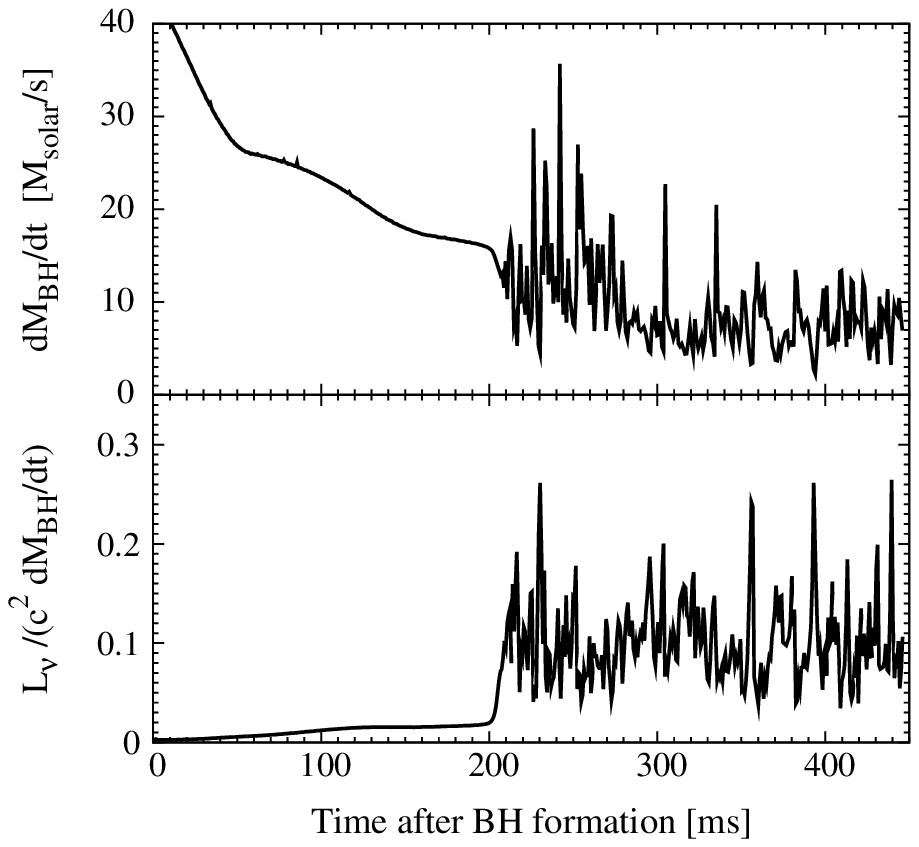}
  \end{center}
  \caption{Mass accretion rate into the black hole $d M_{\rm BH}/dt \equiv \dot{M}_{\rm BH}$ 
    (the upper panel) and an efficiency of neutrino emission $L_{\nu}/\dot{M}_{\rm BH}c^{2}$ 
    (the lower panel) as functions of time after the black hole (BH) formation 
    for the moderately rotating model.
  \label{fig_mdot-r05}}
\end{figure}

\begin{figure}
  \begin{center}
    \epsscale{1.0}
    \plotone{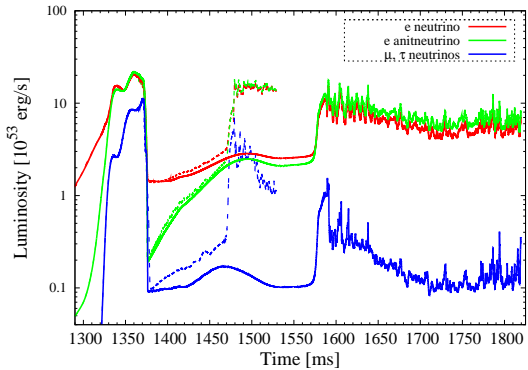}
  \end{center}
  \caption{Time evolution of neutrino luminosities for the moderately rotating model for
    the lower (solid curves) and finer (dashed curves) resolutions.
    A black hole is formed at $t\approx 1373$ ms.
  \label{fig_lum-r05}}
\end{figure}

\begin{figure*}
  \begin{center}
    \epsscale{1.0}
    \plotone{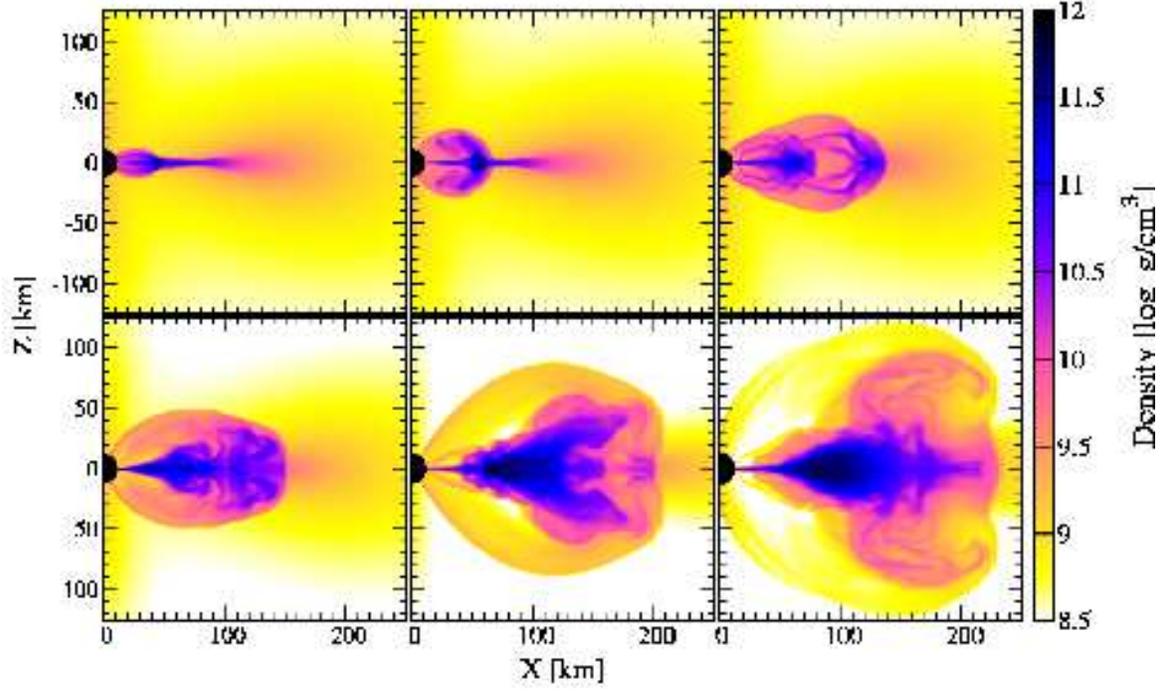}
  \end{center}
  \caption{Contours of rest mass density at $t\approx1578$ (top left), 1584 (top middle), 
           1591 (top right), 1644 (bottom left), 1706 (bottom middle), and 
           1800 ms (bottom right)
           for the moderately rotating model.
  \label{fig_con-rho-r05}}
\end{figure*}

\begin{figure*}
  \begin{center}
    \epsscale{1.0}
    \plotone{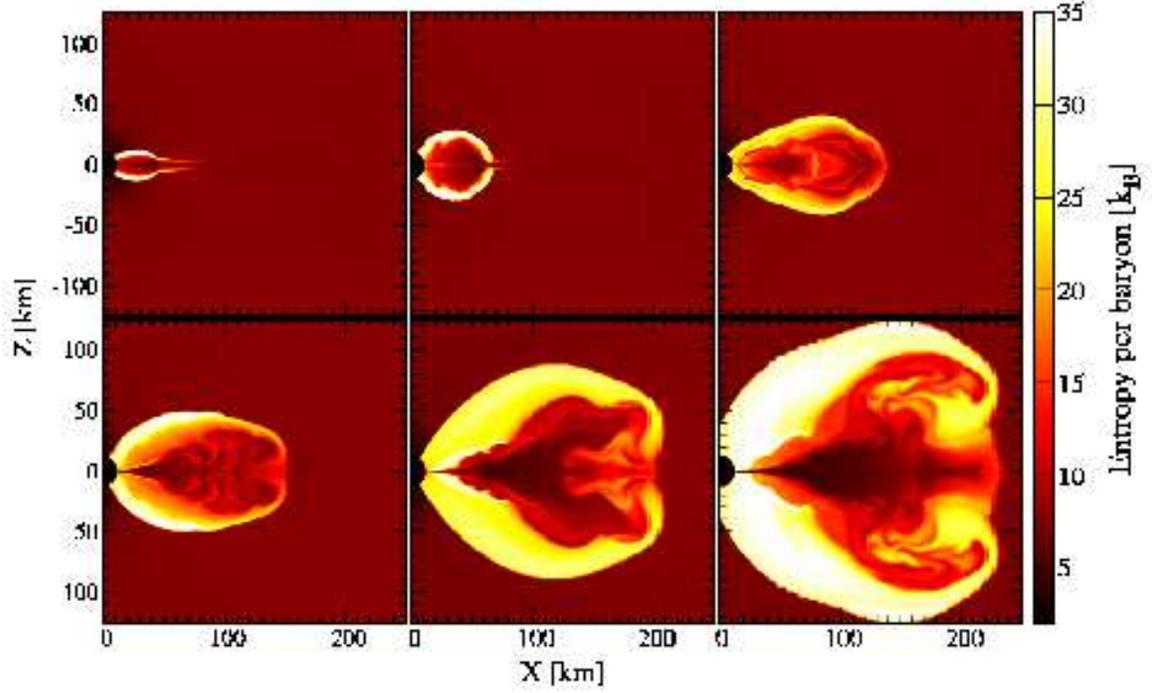}
  \end{center}
  \caption{Contours of entropy per baryon for the moderately rotating model. 
           The selected time slices are the same as those in 
           Figure~\ref{fig_con-rho-r05}.
           \label{fig_con-sen-r05}}
\end{figure*}

\begin{figure*}
  \begin{center}
    \epsscale{1.0}
    \plotone{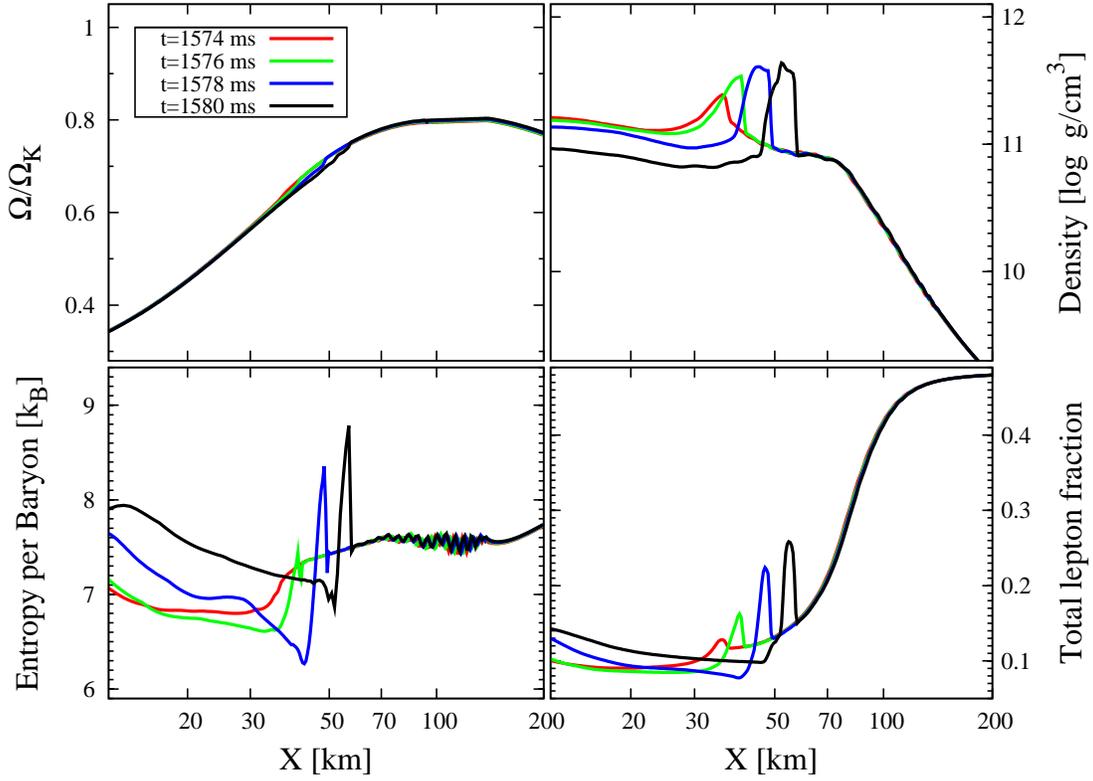}
  \end{center}
  \caption{Profiles of $\Omega/\Omega_{\rm K}$, density, entropy per baryon, 
           and total lepton fraction along the radial direction in the equator
           at $t\approx 1574$, 1576, 1578, and 1580 ms.
           \label{fig_qalx-conv}}
\end{figure*}

\begin{figure*}
  \begin{center}
    \epsscale{1.0}
    \plotone{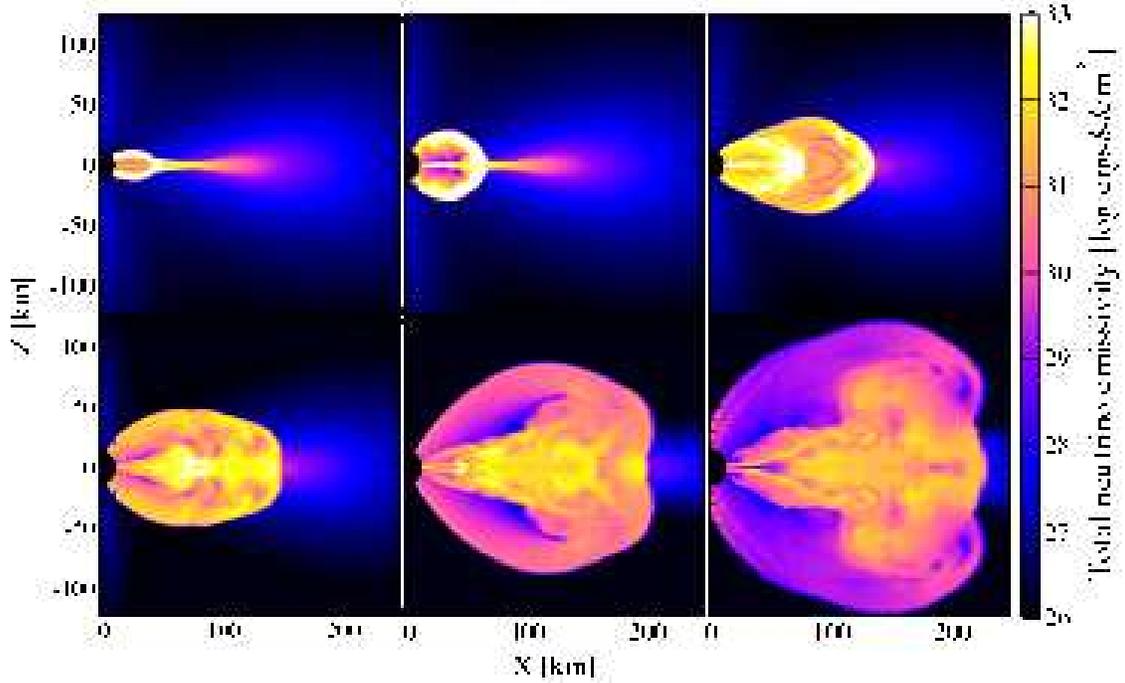}
  \end{center}
  \caption{Contours of the total neutrino emissivity for the moderately rotating model. 
           The selected time slices are the same as those in 
           Figure~\ref{fig_con-rho-r05}.
           \label{fig_con-dq-r05}}
\end{figure*}

\begin{figure*}
  \begin{center}
    \epsscale{1.0}
    \plotone{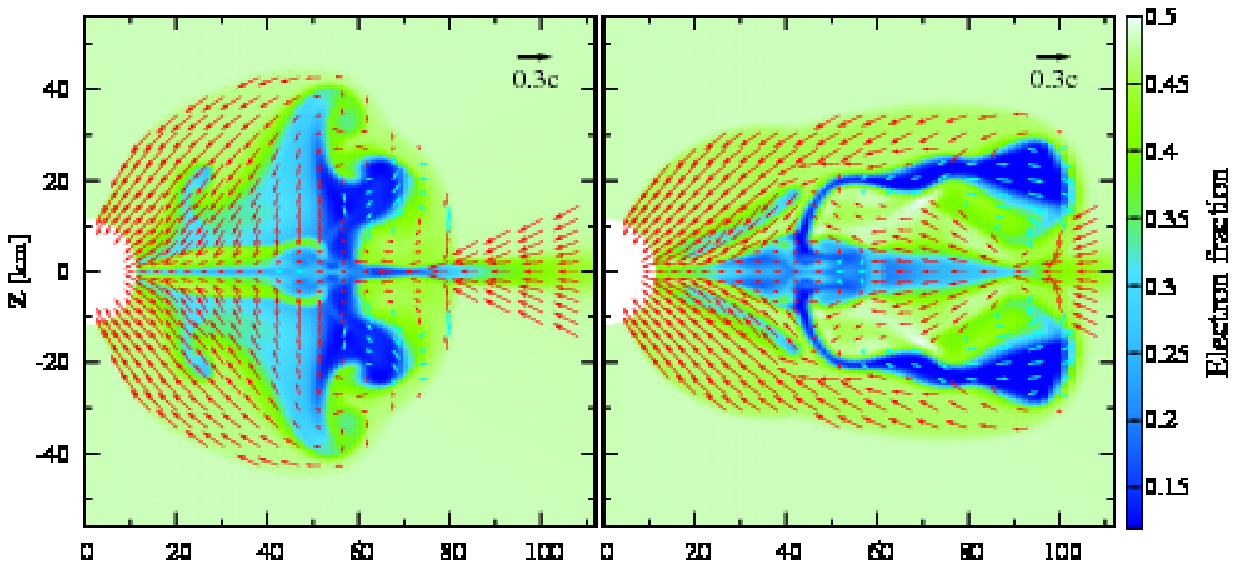}
    \epsscale{1}
    \plotone{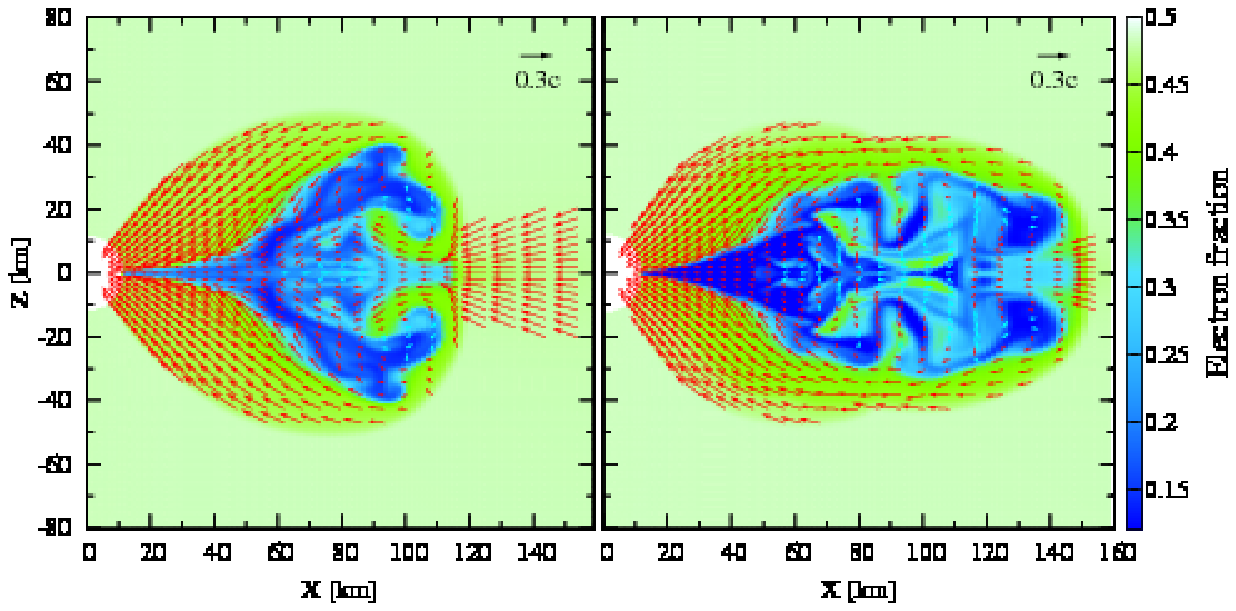}
  \end{center}
  \caption{Contours of electron fraction with velocity fields at $t\approx 1589$ (top left panel),
    1590 (top right panel), 1596 (bottom left panel), and 1644 ms (bottom right panel).
    \label{fig_KH-1}}
\end{figure*}

\begin{figure*}
  \begin{center}
    \epsscale{1.0}
    \plotone{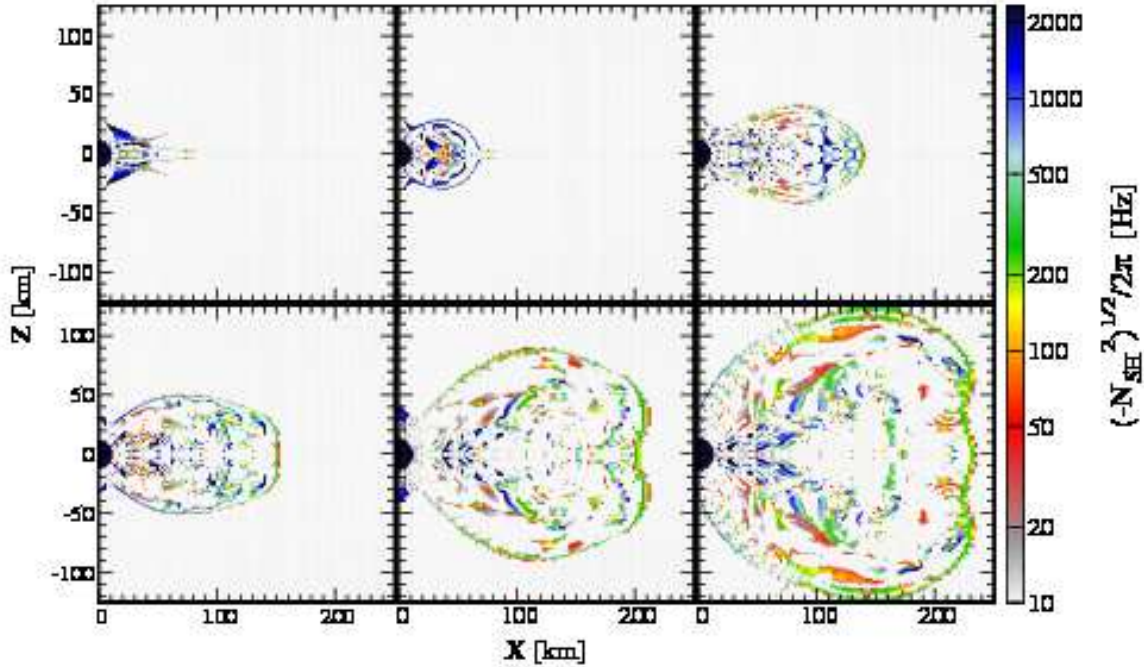}
  \end{center}
  \caption{Contours of the Solberg-Hoiland frequency for the moderately rotating model. 
           The selected time slices are the same as those in 
           Figure~\ref{fig_con-rho-r05}.
           \label{fig_con-BV}}
\end{figure*}

\begin{figure}
  \begin{center}
    \epsscale{1.0}
    \plotone{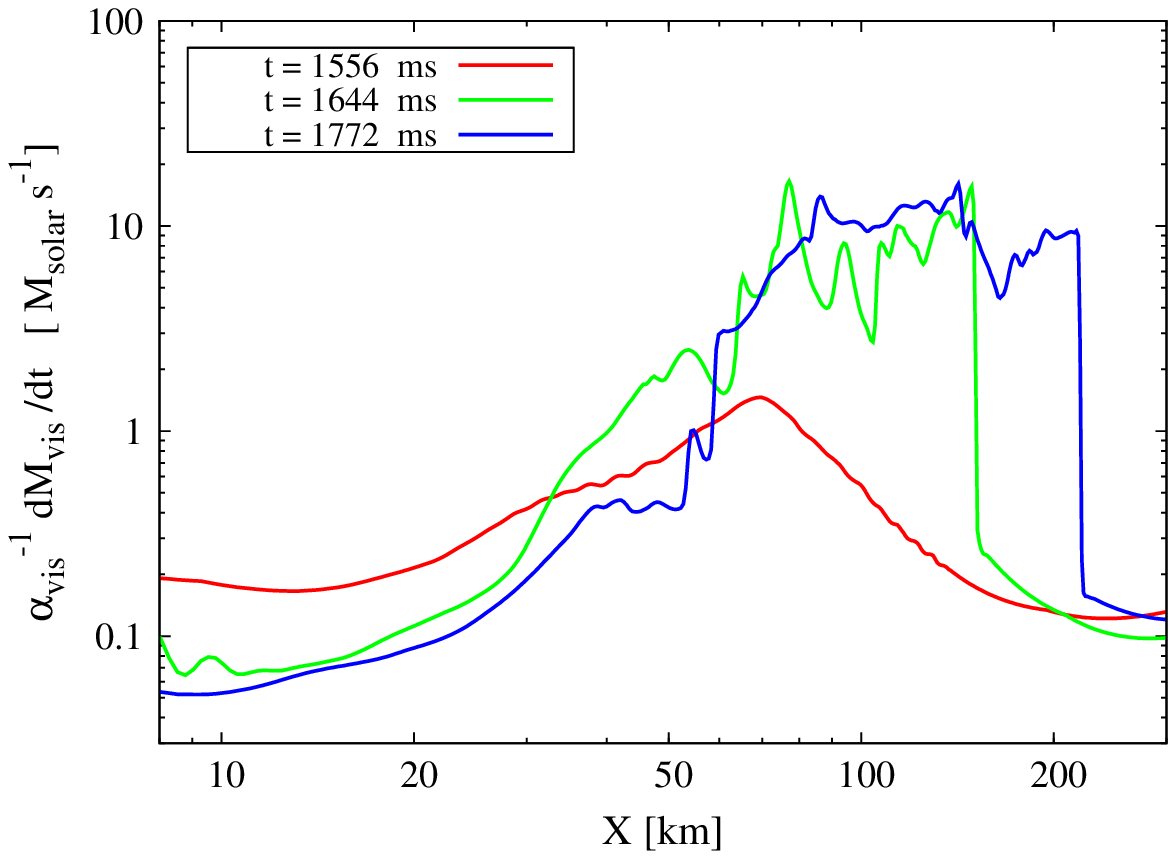}
  \end{center}
  \caption{Profiles of $\dot{M}_{\rm vis}$ along the radial direction in the equator
           in the geometrically-thin-disk phase (at $t\approx 1556$ ms) and  
           the convective phase (at $t\approx 1644$ and $1574$ ms).
           \label{fig_mvis-r05}}
\end{figure}

The basic features of rotational core collapse until the black hole formation 
are qualitatively the same as those of the spherical model:
Gravitational collapse is triggered primarily by the photo-dissociation 
of heavy nuclei; the gas-pressure-dominated bounce occurs at a subnuclear density;
a weak shock wave is formed at the bounce and is stalled quickly; a black hole
is formed soon after the bounce in $\approx 30$--50 ms.
After the black hole formation, on the other hand, the dynamics of infalling material 
is modified by the centrifugal force;
an accretion disk is formed around the black hole as the material with 
sufficient specific angular momentum falls into the central region.
We first describe the feature of the collapse for the moderately rotating model 
in Sections \ref{Sec_disk}, \ref{Sec_torus} and \ref{Sec_conv}.
Then, we discuss dependence of the dynamics of the accretion disk formation
and properties of the disk on the amount of rotation in Section \ref{Sec_rot}.
It is found that the process of the accretion disk formation and properties of 
the disk depend sensitively on the amount of rotation initially given.

\subsubsection{Black hole and thin accretion disk formation} \label{Sec_disk}

In this subsection, we describe features of dynamics of the first $\sim$ 200 ms
after the black hole formation.
We note that time duration of this phase depends on the grid resolution 
but the evolution process does not depend qualitatively on it.
Figure \ref{fig_con-disk} plots contours of density, electron fraction,
entropy per baryon, and temperature at selected time slices around black hole 
formation epoch. 
As in the collapse for the spherical model, the weak bounce occurs
at $t\approx 1339$ ms, and then, convectively unstable regions with negative gradients 
of electron fraction appear when the shock wave is stalled. 
However, because the core immediately collapses to a black hole, the 
convection is only weakly activated and plays a minor role 
(see the left and middle panels in Figure \ref{fig_con-disk}). 
Accompanied with the black hole formation, a geometrically thin, 'sub-Keplerian' disk 
is formed around the black hole (see below for details).
Note that the disk is geometrically thin not due to the neutrino cooling 
(because the disk is optically thick), but mainly due to the ram pressure of infalling 
material (see Eq. \ref{Eq_H-1} and the discussion below).

Figure \ref{fig_BH-r05} plots the time evolution of mass and spin parameter of the 
black hole as well as disk mass ($M_{\rm disk}$). 
At $t\approx 1373$ ms, a black hole of $M_{\rm BH} \approx 6.5M_{\odot}$
with spin parameter of $q_{\rm BH} \approx 0.6$ is formed. 
The initial mass of the black hole is larger than that in the spherical collapse 
because the threshold mass for the black hole formation is larger 
due to effect of the rotation (the centrifugal force).
Note that $M_{ce}$ seems to be a good indicator of mass of a black hole even 
in the presence of a massive accretion disk as suggested in \citet{Shibata07}, 
because the time evolution of $M_{\rm con}$ and $M_{ce}$ approximately agrees
with each other.
The upper panel in Figure \ref{fig_mdot-r05} plots the time evolution of mass 
accretion rate into the black hole ($\dot{M}_{\rm BH}$).
The mass accretion rate soon (10 ms) after the black hole formation is high as 
$\dot{M}_{\rm BH}\approx 40 M_{\odot}$ s$^{-1}$.
The mass accretion rate decreases gradually with time, but even at $t \sim 1800$ ms,
it is still as high as $\dot{M}_{\rm BH} \sim 5$--$10M_{\odot}$ s$^{-1}$ 
(see the upper panel in Figure \ref{fig_mdot-r05}).

Figure \ref{fig_lum-r05} plots the time evolution of neutrino luminosities
for the moderately rotating model. As in the spherical 
model, electron neutrinos are dominantly emitted before the weak bounce,
and electron neutrinos and electron anti-neutrinos are approximately
identically emitted after the bounce.
The luminosity curves of electron neutrinos ($\approx 1.6 \times 10^{54}$ erg/s) 
and anti-neutrinos ($1.4 \times 10^{54}$ erg/s) achieve the first peak soon after the 
weak bounce (at $t \approx 1330$ ms). 
By contrast with the spherical model, the second peak appears
in the neutrino luminosity curves at $t\approx 1360$ ms.
Because oblate (or torus-like) neutrino 'sphere' is formed after the bounce due to 
the rotation, the optical depth of neutrinos is smaller in the $z$-direction
(see the middle panels in Figure \ref{fig_con-disk}).
As a result, neutrinos are more efficiently emitted in the $z$-direction and
this effect constitutes the second peak.
In this phase, more electron anti-neutrinos are emitted than electron neutrinos
($Q^{\rm ec}_{\nu_{e}} \lesssim Q^{\rm pc}_{\bar{\nu}_{e}}$)
because the electrons inside the torus is only weakly degenerate 
$\eta_{e} \sim 1$ due to high temperature and the fraction of neutrons is
larger than that of protons as $X_{p}/X_{n} \sim 0.2$, enhancing the reaction of
$n+e^{+} \rightarrow p+\bar{\nu}_{e}$.

Soon after the black hole is formed, most of the material inside the 
oblate structure is quickly swallowed by the black hole because they 
do not have enough angular momentum to retain in the orbit around the formed black hole.
However, a small amount of the material with sufficient angular momentum forms
a geometrically thin accretion disk around the black hole 
(see the right panels in Figure \ref{fig_con-disk}).
Mass of the geometrically thin disk just after the black hole formation is 
$M_{\rm disk} \approx 0.2 M_{\odot}$ and subsequently decreases to be $\approx 0.1M_{\odot}$
(see the bottom panel in Figure \ref{fig_BH-r05})
because material with high density located near the rotational axis,
which does not have sufficient angular momentum and does not constitute the disk, 
is swallowed by the black hole. 
Then the thin-disk mass relaxes to a quasi-stationary value of $\sim 0.1M_{\odot}$ 
and the net mass infall rate onto the thin disk vanishes approximately 
($\dot{M}_{\rm disk} \sim 0$).
(For sudden increase of $M_{\rm disk}$ at $t\approx 1580$ ms, 
see Section \ref{Sec_torus}.)

The rest mass density and temperature of the thin disk are 
initially $\sim 10^{11}$ g/cm$^{3}$ and $\sim 8$ MeV (see Figure \ref{fig_con-disk}),
and accordingly, the thin disk is optically thick to neutrinos with the 
maximum optical depth of $\tau_{\nu} \sim 4$ (which increases as the material with
high angular momentum falls onto the thin disk).
At the same time, shocks are formed in the inner part of the thin disk, 
converting kinetic energy of infalling materials into thermal energy
(see \citet{Lee06} for discussion of a similar phenomenon).

The shock is successively formed due to infall of the material 
with angular momentum not large enough to retain 
the orbit around the black hole.
After hitting the surface in the inner region of the disk, 
such material falls into the 
black hole quickly because of its insufficient specific angular momentum,
and contributes to a rapid growth of the black hole.
A part of the thermal energy generated at the shock is advected 
together into the black hole (see discussion below).

The thermal energy is also carried away by neutrinos
because the cooling timescale of neutrino emission, $t_{\rm cool}$, 
is short due to the low density and small pressure scale height 
of the disk, $H$ (although the optical depth is greater than unity), as
\beq
t_{\rm cool} \sim \frac{H\tau_{\nu}}{c} \approx 0.12 \, 
\left(\frac{H}{10\, {\rm km}}\right)
\left(\frac{\tau_{\nu}}{4}\right) \, {\rm ms}.
\label{Eq_t_cool}
\eeq
This is much shorter than the advection time scale approximately given by
\beq
t_{\rm adv} \sim \frac{R_{\rm disk}}{v_{\rm adv}} \approx 1.7\, 
\left( \frac{R_{\rm disk}}{50 {\rm km}} \right)
\left( \frac{v_{\rm adv}}{0.1 c} \right)^{-1} \, {\rm ms},
\eeq
where $R_{\rm disk} (\approx r_{\rm ISCO} \gg H)$ and $v_{\rm adv}$ are 
characteristic radius of the disk and characteristic advection velocity.
 
The pressure scale height may be approximately determined by the 
following force balance relation \citep{Sekiguchi07}:
\beq 
{P_{\rm disk}-P_{\rm ram} \over H} \sim 
\frac{GM_{\rm BH}\rho_{\rm disk}H}{R_{\rm disk}^{\ 3}}, 
\label{Eq_H-1}
\eeq
where $\rho_{\rm disk}$ and $P_{\rm disk}$ are characteristic 
density and pressure of the disk, and $P_{\rm ram}$ is the ram 
pressure of the infalling material, respectively. 
Equation (\ref{Eq_H-1}) gives 
\beq
\frac{H}{R_{\rm disk}} \sim 
\left(\frac{P_{\rm disk}-P_{\rm ram}}{10^{30}\ {\rm dyn/cm}^{2}}\right)^{1/2}
\,\left(\frac{\rho_{\rm disk}}{10^{10}\ {\rm g/cm}^{3}}\right)^{1/2} .
\eeq
Because the density and temperature retain to be low due to rapid advection 
and copious neutrino emission, $P_{\rm disk}\sim 10^{30}$ dyn/cm$^{2}$ 
is as small as the ram pressure, approximately written as
\beq
P_{\rm ram} \sim \rho_{\rm f}v_{\rm f}^{2} \sim 10^{30}\,
\left(\frac{\rho_{\rm f}}{10^{10}\,{\rm g/cm}^{3}}\right)
\ {\rm dyn/cm}^{2} ,
\eeq
where
$\rho_{\rm f}$ and $v_{\rm f}\sim (2GM_{\rm BH}/R_{\rm disk})^{1/2} 
\sim 0.4$--0.5$c$ are the density and velocity of the infalling material,
respectively.
Since $|P_{\rm disk}-P_{\rm ram}| \ll P_{\rm disk}$, 
the pressure scale height is very small as $H/R_{\rm disk} \ll 1$
in the early stage of the thin disk.

The lower panel in Figure \ref{fig_mdot-r05} plots an efficiency of neutrino 
emission defined by $L_{\nu, {\rm tot}}/(\dot{M}_{\rm BH}c^{2})$,
where $L_{\nu, {\rm tot}}$ is the total neutrino luminosity.
The efficiency is low as $\sim 0.01$ in the thin accretion disk phase
(until $\approx 200$ ms after the black hole formation).
On the other hand, order of magnitude of the thermal energy generated at 
the shock in the inner region of the thin disk is estimated to give
\beqn
\frac{GM_{\rm BH}\dot{M}}{r} &\sim& 0.1 \dot{M}_{\rm BH}c^{2} \nonumber \\
&\sim& 5 \times 10^{54} 
\left( \frac{\dot{M}_{\rm BH}}{30M_{\odot}~{\rm s}^{-1}} \right)
\, {\rm erg/s},
\eeqn
where $r\sim 0.1 GM_{\rm BH}/c^{2}$ is the distance from the black hole.
Here, it is assumed that most of the material falling onto the system 
experiences the shock heating 
(i.e., the total mass accretion rate $\dot{M}$ is used) and an 
approximation of 
$\dot{M} = \dot{M}_{\rm BH} + \dot{M}_{\rm disk} \approx \dot{M}_{\rm BH}$
is used.

Thus, the neutrino luminosity is by one order of magnitude smaller than
that the energy generated at the shock.
This indicates that the amount of material which experiences the shock 
heating is much smaller than that swallowed into the black hole because of
small geometrical cross section with the disk.
As the material with high specific angular momentum falls onto the disk and 
the size of the disk increases, neutrino luminosities and the shock heating 
efficiency increase (see Figure \ref{fig_lum-r05} and the lower panel 
in Figure \ref{fig_mdot-r05}). 
(For decrease of neutrino luminosities at $t\approx 1470$ ms, 
see Section \ref{Sec_torus}.) 

Because the total mass of the material surrounding the black hole is much
larger than that in the spherical model, the neutrino luminosity remains
high $> 10^{53}$ erg/s even after the black hole formation 
(see Figure \ref{fig_lum-r05}).
For $\sim 200$ ms after the thin disk formation, the luminosity slightly increases but
is kept to be $\sim 2 \times 10^{53}$ erg/s.
Because the duration of the neutrino emission from the thin disk is
much longer than that before the black hole formation, 
neutrinos are likely to be primarily emitted from the accretion disk (torus),
not during the black hole formation, in the moderately rotating model.

\subsubsection{Disk expansion and torus formation} \label{Sec_torus}

Figures \ref{fig_con-rho-r05} and \ref{fig_con-sen-r05} plot 
contours of density and entropy per baryon at selected time slices 
$\sim 200$--400 ms after the black hole formation. 
It is found that the geometrically thin accretion disk formed in the early 
stage expands to form a geometrically thick accretion torus. 
Note that the disk is also 'sub-Keplerain' in this stage (see the top-left panel in
Figure \ref{fig_qalx-conv}).  
The feature of dynamics can be explained as follows.

As the material with higher specific angular momentum in the outer 
region falls onto the disk, the density and mass of the disk increases
(see the bottom panel in Figure \ref{fig_BH-r05}).
This situation is different from that in the early evolution of the geometrically 
thin disk, in which the material with small specific angular momentum dominantly falls.
As a result, neutrino optical depth increases and neutrino cooling timescale 
becomes longer (cf. Eq. (\ref{Eq_t_cool})).
This helps further storing thermal energy inside the disk and the pressure scale 
height increases (see the top-left panel in Figure \ref{fig_con-sen-r05}).

As the thermal energy is stored, the disk height $H$ increases
according to Eq. (\ref{Eq_H-1}).
The density and the temperature ($T_{\rm disk}$)
inside the disk eventually increase to be $\gtrsim 10^{11}$ g/cm$^{3}$ 
and $\gtrsim 10$ MeV (and hence, $P_{\rm disk} \gtrsim 10^{30}$ dyn/cm$^{2}$).
At the same time, the ram pressure decreases to be 
$\lesssim 0.1P_{\rm disk}$ ($\ll P_{\rm disk}$) because the density of 
the infalling material decreases to $\lesssim 10^{9}$ g/cm$^{3}$.
Consequently, $H$ increases to be $\sim R_{\rm disk}$ 
(see the top-middle panels in Figures 
\ref{fig_con-rho-r05} and  \ref{fig_con-sen-r05}).
For $H\gtrsim R_{\rm disk}$, the approximate force 
balance relation (\ref{Eq_H-1}) changes to
\beq 
(P_{\rm disk}-P_{\rm ram})\sim \frac{GM_{\rm BH}\rho_{\rm disk}}{H}.
\eeq
Because the binding due to the gravitational force by the black hole 
decreases as $H$ increases, the disk expands forming a shock wave
once the condition $H \gtrsim R_{\rm disk}$ is satisfied \citep{Sekiguchi07}.
Figure \ref{fig_qalx-conv} shows that the shock is formed at $t\approx 1576$ ms.

The neutrino opacities decrease as the disk expands 
(density and temperature decrease), and accordingly, 
the cooling timescale becomes shorter.
Then, the shock wave is stalled and the disk relaxes to a new geometrically
thick state. The shock becomes a standing accretion shock and 
expands gradually because the material with higher specific angular 
momentum continuously falls onto the shock and also because the 
ram pressure of the infalling material 
continues to decrease (see the bottom panels in Figures 
\ref{fig_con-rho-r05} and \ref{fig_con-sen-r05}).

Note that when the pressure scale height, and thus, the optical depth 
become sufficiently large, 
the neutrino-cooling timescale becomes longer than the advection 
timescale into black hole, and consequently, neutrinos are {\it trapped} 
in the accretion flow.
This can be seen in the time evolution of neutrino luminosities plotted 
in Figure \ref{fig_lum-r05}: At $t\approx 1490$ ms, neutrino 
luminosities start decreasing slightly. 
The trapping of neutrinos are also found in a steady 
high-density accretion disk model \citep{DiMatteo02,Chen07}.
Note also that similar decrease of neutrino luminosities has been found
in the simulations of ordinary core collapse soon after the
onset of neutrino trapping \citep[e.g.,][]{Liebendorfer01}.

Figure \ref{fig_con-dq-r05} plots contours of the total neutrino emissivity 
at selected time slices $\sim 200$--400 ms after the black hole formation. 
Neutrino luminosities are significantly enhanced after the thick torus formation.
The reason for this is mainly that the amount of material which experiences the shock
heating increases. The disk is optically thick to neutrinos at first and becomes optically 
thin as the disk expands. Then, neutrinos trapped inside the torus are emitted.
This feature is somewhat similar to the so-called 'neutrino burst' associated with 
the early shock formation in the ordinary supernova explosion.

After the expansion, the total luminosity reaches $\approx 2 \times 10^{54}$ erg/s 
because amount of material which experiences the shock heating significantly increases. 
Then the efficiency of neutrino emission is as high as
$L_{\nu, {\rm tot}}/(\dot{M}_{\rm BH}c^{2}) \sim 0.1$ 
(see the lower panel in Figure \ref{fig_mdot-r05}).
These agree approximately with the generation rate of thermal energy by 
infalling material on the standing shock,
\beqn
\frac{GM_{\rm BH}\dot{M}}{r} &\sim& 0.1 \dot{M}c^{2} \nonumber \\
&\sim& 2 \times 10^{54} {\rm erg/s} 
\left( \frac{\dot{M}}{10M_{\odot}~{\rm /s}} \right),
\eeqn
where a characteristic value of 
$\dot{M} = \dot{M}_{\rm BH} + \dot{M}_{\rm disk} \sim 10M_{\odot}$ s$^{-1}$
is adopted (see the bottom panel in Figure \ref{fig_BH-r05} and 
the upper panel in Figure \ref{fig_mdot-r05}).
The high efficiency indicates that neutrino optical depth is not very high for the 
neutrino-emission region and that advection of the thermal energy into the
black hole is not very large in this phase because of the quick neutrino emission.

\subsubsection{Convective activities} \label{Sec_conv}

After the formation of the geometrically thick torus, 
convective motions are excited near the shocked region in the torus.
The origin of the convection is explained as follows.

The shock heating is more efficient in an inner part of the torus
because kinetic energy of infalling material is larger
(see the top-left panel in Figure \ref{fig_con-sen-r05}).
On the other hand, the neutrino cooling is less efficient in the inner part of 
the torus because of its higher density and resulting larger optical depth.
Then, the entropy per baryon becomes higher in the shocked inner region of
the torus (see Figure \ref{fig_con-sen-r05}),
and consequently, regions of negative entropy gradient along the radial direction 
near the equatorial plane are developed.
Also, because neutrinos are trapped and $\beta$-equilibrium is achieved 
in the inner part of the torus, the total lepton fraction increases inward.
These tendencies are enhanced as the accretion of the material with
higher angular momentum proceeds.

The condition for convective instabilities to occur is 
given by the so-called Solberg-Hoiland criterion \citep[e.g.,][]{Tassoul78},
\beq
N_{\rm SH}^{\ 2} = N_{\rm BV}^{\ 2} + \kappa^{2} < 0,
\label{Eq_fq}
\eeq 
where $N_{\rm BV}$ is the Brunt-V\"ais\"al\"a frequency given by
\citep[e.g.,][]{Lattimer81}
\beqn
N_{\rm BV}^{\ 2} &=& \frac{g_{\rm eff}}{\rho}
\left(\frac{\partial \rho}{\partial P}\right)_{s, Y_{l}} \nonumber \\
&& \times 
\left[
 \left(\frac{\partial P}{\partial s}\right)_{\rho, Y_{l}}
 \left(\frac{d s}{d r}\right)
+\left(\frac{\partial P}{\partial Y_{l}}\right)_{\rho, s}
 \left(\frac{d Y_{l}}{d r}\right)
\right] ,
\label{Eq_BV}
\eeqn
and $\kappa$ is the epicyclic frequency which may be written for
nearly circular orbits as \citep[e.g.,][]{Binney87}
\beq
\kappa^{2} = \varpi \frac{d\Omega^{2}}{d \varpi} + 4\Omega^{2}.
\label{Eq_epi}
\eeq 

Figure \ref{fig_qalx-conv} plots the profiles of angular velocity, 
total lepton fraction, and entropy per baryon along the radial 
direction in the equator after the convection sets in. 
It is clearly shown that negative entropy gradient is formed in 
several regions inside the torus, and drives convection (see Figures
\ref{fig_con-rho-r05} and \ref{fig_con-sen-r05}).
Rotation does not play an important role in suppressing the convective activities
because the angular velocity $\Omega$ is smaller than the Kepler 
angular velocity given by
\beq
\Omega_{\rm K} = \left[
\sqrt{\frac{r^{3}}{GM_{\rm BH}}} + q_{\rm BH}\frac{GM_{\rm BH}}{c^{3}}
\right]^{-1},
\eeq
(see the top-left panel in Figure \ref{fig_qalx-conv}), and thus,
Coriolis force is not large enough.

The convective flows cannot move freely because 
the material infalling from the outside of the torus prevents the free expansion of the 
convective components (see the top-middle panel in Figure \ref{fig_con-rho-r05}).
Figure \ref{fig_KH-1} plots contours of electron fraction with velocity 
fields.
Interacting with the thin accretion flows, a part of the convective 
flows is swerved to form finger-like structure 
(see the top-right panel in Figure \ref{fig_KH-1}).
Then, the convective components form a swirl.
Note that regions with velocity shear appear at the interface between the convective 
fingers and the accretion flows (see the right panel in Figure \ref{fig_KH-1}),
and hence, the Kelvin-Helmholtz instability could be developed at the 
interface, generating turbulent motions 
(see the bottom-left panel in Figure \ref{fig_KH-1}).

In addition, oscillations of the standing shock wave are induced.
Such shock oscillations are proposed in a different context 
to explain quasi-periodic oscillations of X-ray binaries \citep{Molteni96} and
found in a recent Newtonian simulation of sub-Keplerian accretion flows around 
a black hole \citep{Giri10}.

Associated with the convective motions, many shock waves are formed
and accretion flows show very complicated features.
Because of interplay of the neutrino-trapping, the Kelvin-Helmholtz instability, 
and the convective shock, the accretion flow remains convectively unstable.
Figure \ref{fig_con-BV} shows the Solberg-Hoiland frequency, $N_{\rm SH}$ defined 
in Eq. (\ref{Eq_fq}). 
The effective gravity appeared in Eq. (\ref{Eq_BV}) is approximately 
evaluated using the Newtonian gravity as
$g_{\rm eff} = GM_{\rm BH}/r^{2}$.
As this figure shows, several regions inside the standing shock remain 
convectively unstable.

As a natural consequence of the convective activities of the accretion flow,
neutrino luminosities vary violently in time (see Figure \ref{fig_lum-r05}).
If GRBs are driven by the pair annihilation of neutrinos and anti-neutrinos,
such time-variability may explain the observed 
time-variability of GRB light curves.
Furthermore, electrons in the convective regions are only weakly degenerate
due to the high entropy and temperature. Consequently, the emissivities of 
electron neutrinos and electron anti-neutrinos are approximately identical 
($Q^{\rm ec}_{\nu_{e}} \sim Q^{\rm pc}_{\bar{\nu}_{e}}$).
This is favorable for the pair annihilation of neutrinos to electron-positron
pairs because its rate is proportional to $L_{\nu}L_{\bar{\nu}}$ 
(see Section \ref{Sec_GRB}).
We finally note that the total energies emitted in neutrinos over the entire time
of the simulations are $E_{\nu, {\rm tot}} \approx 3.8 \times 10^{53}$, 
$3.9 \times 10^{53}$, and $9.4 \times 10^{52}$ erg for electron neutrinos,
electron anti-neutrinos, and {\it total} of $\mu$ and $\tau$ neutrinos.

\subsubsection{Effect of viscosity and formation of viscous accretion disk}

Finally we remark possible effects of viscosity in the evolution of the accretion disk,
which are not taken into account in our simulation.
Assuming that the disk (or torus) can be described by the standard disk model 
with $\alpha$-viscosity 
\citep{Shakura73}, mass accretion rate of disk material into the black hole due to 
the viscous transport of angular momentum ($\dot{M}_{\rm vis}$) is written as
\beq \label{Eq_mdot_vis}
\dot{M}_{\rm vis} \sim 4\pi \alpha_{\rm vis} P H \Omega^{-1},
\eeq
where $\alpha_{\rm vis}$ is the viscous parameter and 
the pressure scale height is approximately estimated by
\beq
H \approx \sqrt{ \frac{P}{\rho} \frac{r^{\ 3}}{GM_{\rm BH}}}\, .
\eeq
Figure \ref{fig_mvis-r05} plots characteristic values of $\dot{M}_{\rm vis}$ 
along the radial direction in the equatorial plane
in the geometrically-thin-disk phase (at $t\approx $ 1556 ms),
early (at $t\approx $ 1644 ms) and late (at $t\approx $ 1772 ms) 
stages of the convective phase. 
During the evolution of the accretion disk, viscosity is not likely to 
play an active role as described in the following. 

In the geometrically-thin-disk phase, the predicted viscous mass accretion rate is
small as $\dot{M}_{\rm vis} \lesssim 0.1 M_{\odot}$ s$^{-1}$ for a relatively 
large viscous parameter of $\alpha_{\rm vis} = 0.1$. Then characteristic timescale
for viscous mass accretion is $\sim 1$ s because the disk mass is 
$M_{\rm disk} \sim 0.1 M_{\odot}$ (see Figure \ref{fig_BH-r05}), which is much
longer than the duration of the geometrically-thin-disk phase $\sim 200$ ms.. 
Thus, the viscosity will not play an important role in the geometrically-thin-disk phase.

In the convective phase, the viscous mass accretion rate becomes large
as $\dot{M}_{\rm vis} \sim M_{\odot}$ s$^{-1}$ for $\alpha_{\rm vis} = 0.1$. 
On the other hand, the mass infall rate onto the torus is 
$\dot{M}_{\rm disk} \sim 3$--$4M_{\odot}$ s$^{-1}$ (see Figure \ref{fig_BH-r05}), 
which is larger than the viscous mass accretion rate.
Thus, effect of viscosity is not likely to play a central role and 
the disk will accumulate mass even in the presence of the viscosity.

The disk will spread outward with accumulating mass until 
the viscous mass accretion rate exceeds the infall mass accretion rate
onto the disk
($\dot{M}_{\rm disk} \sim 4\pi R_{\rm disk}^{2}\rho_{\rm f}v_{\rm f}$).
When $\dot{M}_{\rm disk}$ becomes smaller and the torus becomes more massive
due to accretion of material from outer regions, the viscosity will play
an important role on evolution and dynamics of the torus.
Over the past decade, many groups have studied properties of the 
viscous accretion disk around a black hole
\citep{Popham99,Narayan01,DiMatteo02,Kohri02,Kohri05,Gu06,Chen07,Kawanaka07}.
Such studies have successfully explained the energetics of LGRBs.

It should be note that in the viscous accretion phase, 
the material with low angular momentum will also fall in the vicinity of the
black hole and shock dissipation of the infall kinetic energy will also occur. 
Material with high angular momentum can dissipate their infall 
kinetic energy on the standing shock before they reach the centrifugal barrier.
The amount of such materials depends on the initial density and rotational 
profile yet poorly known. There might be substantial amount of mass accretion 
and energy generation due to such processes.

\subsection{Dependence on grid resolution and numerical accuracy}

\begin{figure}
  \begin{center}
    \epsscale{1.0}
    \plotone{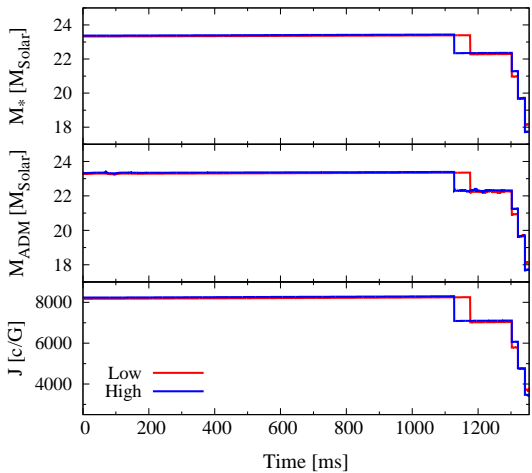}
  \end{center}
  \caption{Time evolution of 
    the total baryon mass (the top panel),
    the total ADM mass (the middle panel), and the total angular 
    momentum (the bottom panel) 
    for the moderately rotating model.
    The red and blue curves correspond to the results in the lower resolution
    and in the higher resolution, respectively.
  \label{fig_conserv-r05}}
\end{figure}

\begin{figure}
  \begin{center}
    \epsscale{1.0}
    \plotone{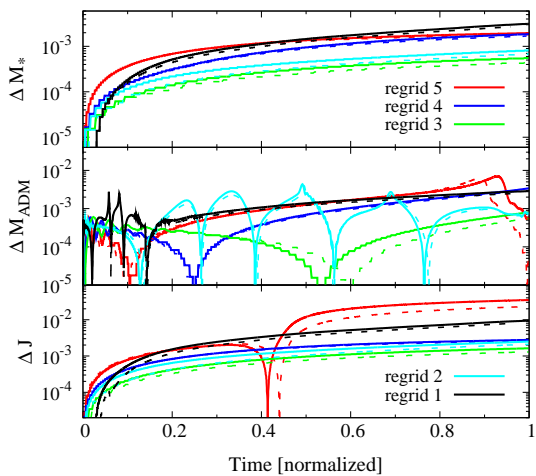}
  \end{center}
  \caption{Time evolution (normalized) in each regrid level of 
    the total baryon mass (the top panel),
    the total ADM mass (the middle panel), and the total angular 
    momentum (the bottom panel) 
    for the moderately rotating model.
    The solid curves correspond to the results in the lower resolution
    and the dashed curves to those in the higher resolution.
  \label{fig_conserv-r05n}}
\end{figure}

\begin{figure}
  \begin{center}
    \epsscale{1.0}
    \plotone{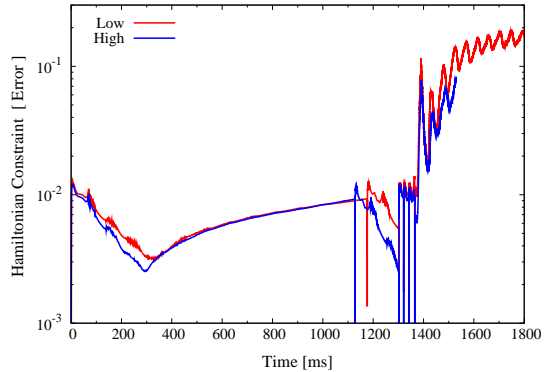}
  \end{center}
  \caption{Time evolution of the Hamiltonian constraint error 
    for the moderately rotating model.
    The red and blue curves correspond to the results in the lower resolution
    and in the higher resolution, respectively.
  \label{fig_const-r05}}
\end{figure}

\begin{figure}
  \begin{center}
    \epsscale{1.0}
    \plotone{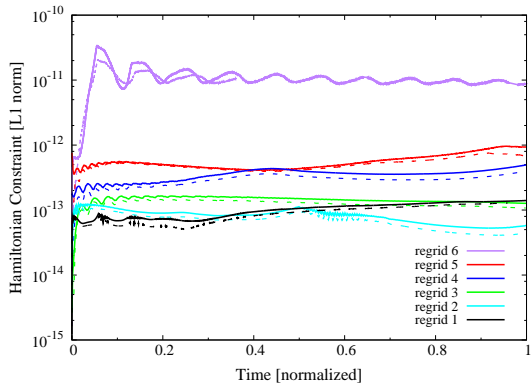}
  \end{center}
  \caption{Time evolution (normalized) in each regrid level of L1 norm of 
    the Hamiltonian constraint for the moderately rotating model.
    The solid curves correspond to the results in the lower resolution
    and the dashed curves to those in the higher resolution.
  \label{fig_const-r05n}}
\end{figure}

Because the present simulation is long-term one, we here describe
dependence of results on the grid resolution and numerical accuracy.
In Figure \ref{fig_lum-r05}, we compare the time evolution of neutrino 
luminosities derived both in the high (dashed curves) and low (solid curves)
resolution runs. The neutrino luminosities in the two grid resolutions 
agree very well until the black hole formation, 
indicating that converged results are obtained for such phase. 
In the geometrically thin disk phase, on the other hand, 
the luminosities in the finer resolution are systematically higher than 
those in the lower resolution. This is because the vertical structure of 
the geometrically thin disk and  shock-heated region are more accurately 
resolved in the finer resolution, and hence, the maximum temperature is 
higher in the finer resolution. 
Also, the geometrically thin disk more quickly expands to be the 
geometrically thick disk. This is because the thermal energy is more efficiently 
stored in the disk because neutrino 
opacities are larger due to the higher density and temperature.
These results indicates the importance of resolving the vertical structure of
the geometrically thin disk for the quantitative study. 
If the grid resolution is not sufficient, 
a geometrically thin disk may remain thin instead of expanding to be thick
torus.

Note that the effects of grid resolution works in a positive manner in 
our results, that is, the transition of a thin disk to a thick disk is
more likely to occur. We therefore safely conclude that qualitative feature of our 
results does not depend on the grid resolution.

To check the accuracy of our results, conservations of
the baryon mass ($M_{*}$), the ADM mass ($M_{\rm ADM}$) \citep[e.g,][]{York79}, 
and the total angular momentum ($J$), as well as violations of the Hamiltonian 
constraint are monitored during the simulation.
Figure \ref{fig_conserv-r05} displays the time evolution of these quantities.
The several discontinuous changes correspond to the regridding procedures where
outer low density region which does not affect the evolution of the central region 
is discarded. In each regridding level, $M_{*}$, $M_{\rm ADM}$, and $J$ conserve well. 
To see this more quantitatively, we display the time evolution of error in each level 
of the regridding until the black hole formation in Figure \ref{fig_conserv-r05}.
The error is given by
\beq
\Delta Q_{{\rm regrid\,}i}(t) = 
\left| \frac{Q_{{\rm regrid\,}i}(t) - Q_{{\rm regrid\,}i}(0)}{Q_{{\rm regrid\,}i}(0)} \right|,
\eeq
where $Q_{{\rm regrid\,} i}$ denotes the conserved quantities $M_{*}$,
$M_{\rm ADM}$, and $J$ in the $i$-th regrid level. 
For the purpose of facilitating visualization, the time is normalized by the 
duration of each regridding level.

The error of conservation of total baryon mass grows monotonically in time,
while it is small as $O(10^{-3})$. A part of the error is caused by the outer 
boundary conditions for fluid quantities where a simple copy is imposed.
The error of the ADM mass shows an oscillating behavior caused by
the regridding procedure, and also is small as $\lesssim 1$\%. 
The error in total angular momentum is also small as a few percent, indicating 
good accuracy of conservation.
Note that after the black hole formation, we start to adopt the 
excision procedure in solving hydrodynamic equations, and consequently,
these quantities do not conserve.

Figure \ref{fig_const-r05} plots the time evolution of the Hamiltonian constraint 
error defined by \cite{Shibata03a}
\beqn
&& {\rm ERROR}={1 \over M_*} \int \rho_* |V| d^3x, \label{H-error}\\
&& V={\displaystyle \tilde \Delta \psi - {\psi \over 8}\tilde R +2\pi E \psi^5
+{\psi^5 \over 8}\tilde A_{ij}\tilde A^{ij}-{\psi^ 5 \over 12}K^2
\over \displaystyle
|\tilde \Delta \psi| + \Big|{\psi \over 8}\tilde R \Big| +2\pi \rho_{h} \psi^5
+{\psi^5 \over 8}\tilde A_{ij}\tilde A^{ij}+{\psi^ 5 \over 12}K^2},
\eeqn
where where $\psi \equiv e^{\phi}$, and
$\tilde \Delta$ denotes the Laplacian with respect to $\tilde \gamma_{ij}$. 
Namely, we use $\rho_*$ as a weight factor for the average.
This weight factor is introduced to monitor whether the main bodies of
the system (inner cores and dense matter regions), 
in which we are interested, are accurately computed or not. 

The several distinct spikes correspond to the regridding procedures where
the Hamiltonian constraint equation is solved numerically.
Until the black hole is formed, the constraint violation is very small as 
$\lesssim 10^{-2}$ and no signal of the increase is seen. 
After the black hole formation, degree of the violation becomes worse
because of the excision procedure. However, the violation is still 
small as $\sim 10^{-1}$, indicating the good accuracy of the simulation.
Note that the integration in Eq. (\ref{H-error}) includes the inside the black hole.
Figure \ref{fig_const-r05n} plots the time evolution (normalized) of the L1 norm of the 
Hamiltonian constraint in each regrid level. Again, the violation does not show 
the signal of rapid increase.

\subsection{Dependence on rotation}\label{Sec_rot}

In this section, we describe dependence of the formation process of the 
black hole and surrounding accretion disk, the convective activities inside the disk, 
and the emissivity of neutrinos, on the degree of initial rotation.

\subsubsection{Slowly rotating model}

\begin{figure}
  \begin{center}
    \epsscale{1.0}
    \plotone{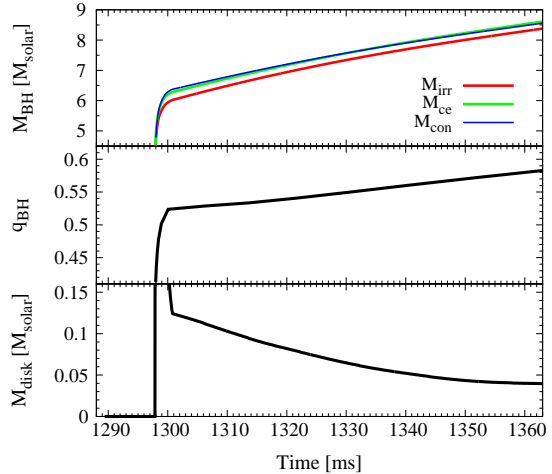}
  \end{center}
  \caption{Time evolution of mass (the top panel) and the non-dimensional 
    spin parameter (the middle panel) of the black hole and disk mass (the bottom panel) 
    for the slowly rotating model.
  \label{fig_BH-r04}}
\end{figure}

\begin{figure}
  \begin{center}
    \epsscale{1.0}
    \plotone{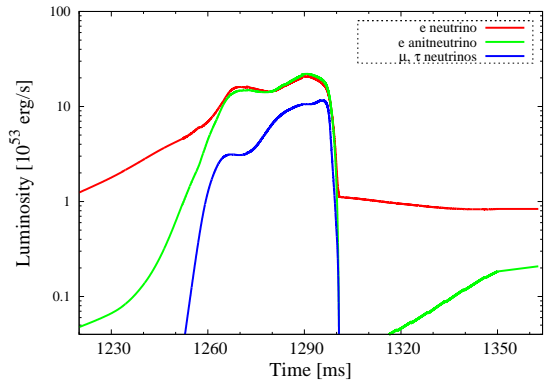}
  \end{center}
  \caption{Time evolution of neutrino luminosities for the slowly rotating model.
  \label{fig_lum-r04}}
\end{figure}

\begin{figure}
  \begin{center}
    \epsscale{1.0}
    \plotone{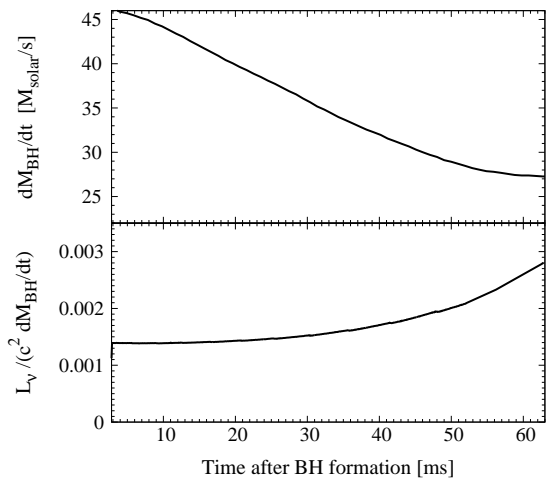}
  \end{center}
  \caption{Mass accretion rate into the black hole $dM_{\rm BH}/dt \equiv \dot{M}_{\rm BH}$ 
    (the upper panel) and efficiency of neutrino emission $L_{\nu}/\dot{M}_{\rm BH}c^{2}$ 
    (the lower panel) as functions of time after the black hole (BH) formation 
    for the slowly rotating model.
  \label{fig_mdot-r04}}
\end{figure}

In the slowly rotating model, a black hole with  $M_{\rm BH} \approx 6.3 M_{\odot}$
and $q_{\rm BH} \approx 0.53$ is formed at $t\approx 1298$ ms.
The mass and spin parameter are only slightly smaller than those in the
moderately rotating model.
Figure \ref{fig_BH-r04} plots the time evolution of mass and spin parameter of the 
black hole as well as disk mass.
The mass accretion rate into the black hole soon (10 ms) after the black hole formation 
is $\dot{M}_{\rm BH}\approx 45 M_{\odot}$ s$^{-1}$ 
(see the upper panel in Figure \ref{fig_mdot-r04}), which is slightly larger than 
that in the moderately rotating model.
The spin parameter remains modest but gradually increases as in the moderately rotating
model.

As in the collapse of the moderately rotating model, a geometrically thin 
(but optically thick) accretion disk is formed soon after the black hole formation. 
In this case, a fraction of the material that forms the disk is smaller than
that for the moderately rotating model due to lower specific angular 
momentum of fluid elements in the slowly rotating model, and hence, 
the disk mass is smaller as $M_{\rm disk}\sim 0.05 M_{\odot}$ 
than that in the moderately rotating model and $\dot{M}_{\rm disk} < 0$ 
(see the bottom panel in Figure \ref{fig_BH-r04}).
However, a part of the material that falls onto the disk still produces shock waves
in the inner part of the disk. 
Thermal energy generated at the shock is not efficiently stored in the disk in the
early stage because most of the shocked material is advected into the black hole and 
neutrinos carry away thermal energy.
Then, the disk remains geometrically thin for a long time (at least $\gtrsim 100$ ms)
after the formation of the black hole.

Figure \ref{fig_lum-r04} plots the time evolution of neutrino luminosities.
Before the black hole formation, the luminosity curves are similar to those
in the moderately rotating model.
It is found that the geometrically thin accretion disk emits $\approx 10^{53}$ erg/s
by neutrinos. This magnitude is by factor of $\sim 2$ smaller than that for the
moderately rotating model. 
The efficiency of neutrino emission is 
$L_{\nu, {\rm tot}}/(\dot{M}_{\rm BH}c^{2}) \approx 0.002$--0.003, which is 
by factor of $\sim 3$ smaller than that for the moderately rotating model 
(see the lower panel in Figure \ref{fig_mdot-r04}), indicating that less
amount of material experiences the shock heating, and that more thermal energy 
is advected into the black hole before released by neutrinos
due to slower rotation and resulting shorter advection timescale.

We do not find any enhancement of neutrino luminosity after 
the black hole formation in our simulation time. 
However, after the free-fall timescale of $\sim$seconds, the material with
higher specific angular momentum may eventually form a dense disk.
Then, thermal energy may be stored inside the disk, and 
the disk may expand to be a geometrically thick torus when the ram pressure of
the infalling material becomes sufficiently small.
Furthermore, provided that the total mass accretion rate 
is sufficiently high as $\dot{M} \gtrsim M_{\odot}$ s$^{-1}$, neutrinos will 
be trapped in the inner region of the disk, and convective activities may set in as 
in the moderately rotation model (see discussion in Section \ref{Sec_spin}). 
If so, it is expected that neutrino luminosities are enhanced and show 
rapid time-variability as in the moderately rotating model.

\subsubsection{Rapidly rotating model}

\begin{figure}
  \begin{center}
    \epsscale{1.0}
    \plotone{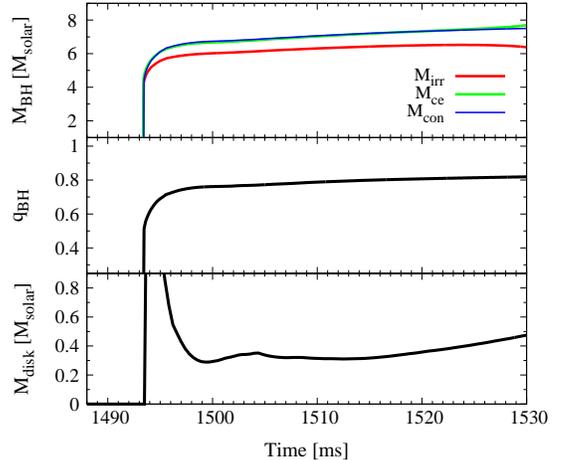}
  \end{center}
  \caption{Time evolution of mass (the top panel) and the non-dimensional 
    spin parameter (the lower panel) of the black hole and disk mass (the bottom panel) 
    for the rapidly rotating model.
  \label{fig_BH-r06}}
\end{figure}

\begin{figure*}
  \begin{center}
    \epsscale{1.0}
    \plotone{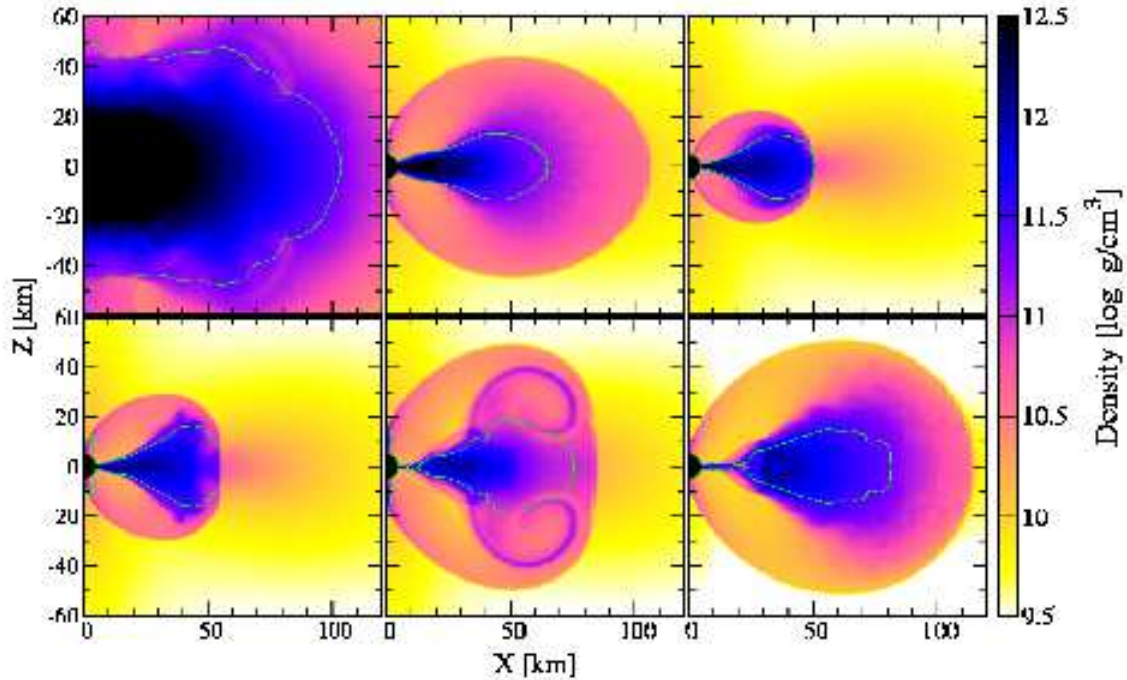}
  \end{center}
  \caption{Contours of rest mass density at $t \approx1495$ (top left), 1497 (top middle), 
           1499 (top right), 1500 (bottom left), 1502 (bottom middle), and 1535 ms 
           (bottom right) for the rapidly rotating model. The green curves indicate
           the region where $\tau_{\nu} = 5$ for electron neutrinos.
           \label{fig_con-rho-r06}}
\end{figure*}

\begin{figure}
  \begin{center}
    \epsscale{1.0}
    \plotone{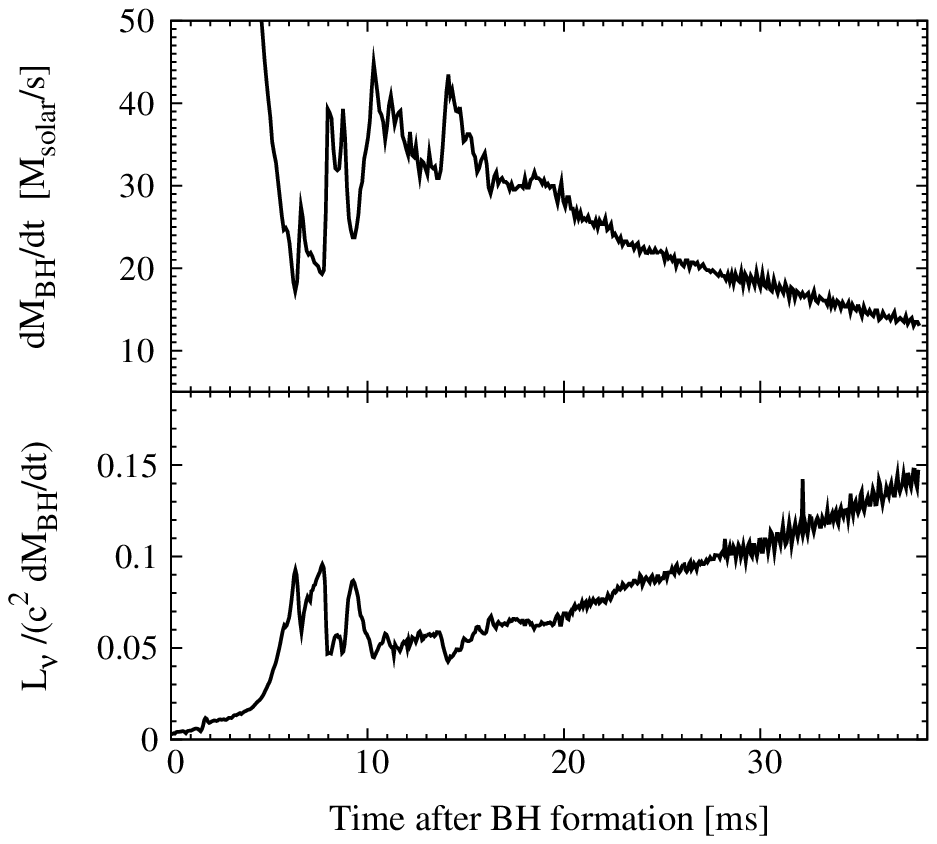}
  \end{center}
  \caption{Mass accretion rate into the black hole $dM_{\rm BH}/dt \equiv \dot{M}_{\rm BH}$ 
    (the upper panel) and efficiency of neutrino emission $L_{\nu}/\dot{M}_{\rm BH}c^{2}$ 
    (the lower panel) as functions of time after the black hole (BH) formation 
    for the rapidly rotating model.
  \label{fig_mdot-r06}}
\end{figure}

\begin{figure}
  \begin{center}
    \epsscale{1.0}
    \plotone{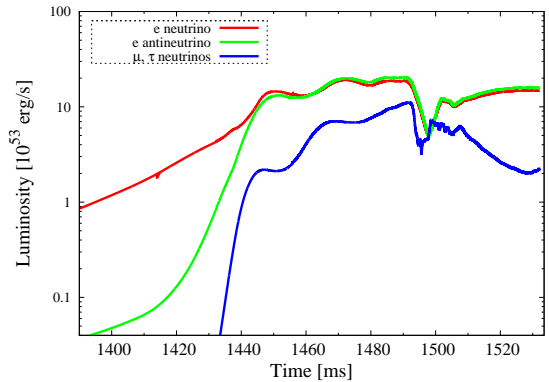}
  \end{center}
  \caption{Time evolution of neutrino luminosities for the rapidly rotating model.
    Note that the black hole is formed at $t\approx 1494$ ms.
    \label{fig_lum-r06}}
\end{figure}

\begin{figure*}
  \begin{center}
    \epsscale{1.0}
    \plotone{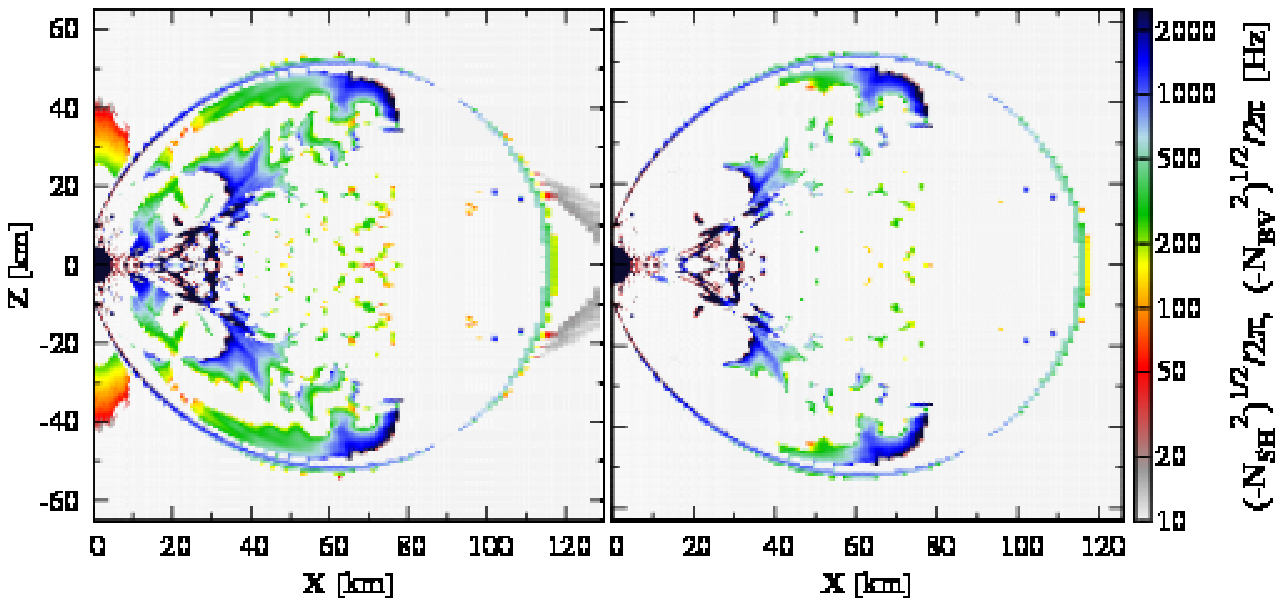}
  \end{center}
  \caption{Contours of the Brunt-V\"ais\"al\"a frequency (left panel) and 
    the Solberg-Hoiland frequency defined by Eq. (\ref{Eq_fq}) (right panel) at 
    $t \approx1534$ ms for the rapidly rotating model. 
    \label{fig_con-BV-r06}}
\end{figure*}

In the rapidly rotating model, a black hole is first formed at $t \approx 1494$ ms 
with mass of $\approx 6.8 M_{\odot}$ and the non-dimensional spin parameter of 
$\approx 0.8$.
Figure \ref{fig_BH-r06} plots the time evolution of mass and spin parameter of
the black hole together with disk mass.
The spin parameter is much larger than that in the moderately rotating
model as expected form Figure \ref{fig_jISCO}. 

In the rapidly rotating model, the disk formation process is 
qualitatively different from that in the moderately rotating model. 
Figure \ref{fig_con-rho-r06} plots contours of rest mass density at 
selected time slices.
The contour curve of $\tau_{\nu_{e}} =5$ is shown together as an approximate
boundary of occurrence of the neutrino trapping.
Inside this curve, neutrinos are trapped because 
$t_{\rm adv} (\sim R_{\rm disk}/v_{\rm adv} \sim R_{\rm disk}/0.1c) 
\sim t_{\rm cool} (\sim H\tau_{\nu}/c)$ for $R_{\rm disk} \sim 2H$.

In the moderately rotating model, a geometrically thin accretion disk is first formed, 
and then, it expands to be a geometrically thick torus. 
In the rapidly rotating model, by contrast, a geometrically thick torus is 
formed immediately after the black hole formation because the pressure gradient and 
the angular momentum of the fluid near the equator are large enough that it retains 
an orbit outside the ISCO. The disk at this phase is still 'sub-Keplerian' with 
$\Omega/\Omega_{K} \approx 0.8$ at its maximum and the pressure gradient plays a 
role in the immediate torus formation.
Reflecting the torus formation, $M_{\rm disk}$ is much
larger as $\approx 0.4 M_{\odot}$ than that in the slowly and moderately rotating models
(see the bottom panel in Figure \ref{fig_BH-r06}).  
Shock waves formed at the weak bounce are not swallowed into the black hole and a 
torus-shaped standing accretion shock remains around the black hole. 

Associated with the torus formation, the mass accretion rate into the black hole 
just after the black hole formation shows non-monotonic behavior by contrast with 
the slowly and rapidly rotating models 
(see the upper panel in Figure \ref{fig_mdot-r06}).
The mass accretion rate quickly drops to be $\dot{M}_{\rm BH} \approx 20 M_{\odot}$ s$^{-1}$
at $t\approx 6$ ms after the black hole formation because of the 
centrifugal and pressure-supported hangup of the torus. 
The subsequent oscillating behavior is due to mass accretion associated with the 
oscillation of the torus.
Then the mass accretion rate decreases quickly with time because the centrifugal 
force of the infalling material prevents the rapid accretion into the black hole.
Note that the pressure gradient plays a role also in this phase.
The mass accretion is expected to cease when $M_{\rm BH} \approx 12 M_{\odot}$.

Figure \ref{fig_lum-r06} plots the neutrino luminosities as a function of time. 
Until the onset of the weak bounce (until the first local peak), 
the luminosity curves are similar to those in other models. 
After the weak bounce occurs, the material near the rotation 
axis starts collapsing, and as a result, the temperature increases 
due to compression and the optical depth near the rotation axis 
relatively decreases. 
Then, second local peak (at $t \approx 1475$ ms) associated with 
a substantial emission from the vicinity of rotation axis appears. 
This is the same feature as found in the slowly and moderately 
rotating models. In the rapidly rotating model, in addition, 
third local peak appears just before black hole formation at 
$t \approx 1494$ ms. This is due to the fact that a dense 
torus, which subsequently falls into the black hole, is  
formed (see the first panel in Figure \ref{fig_con-rho-r06}) 
and emits a large amount of neutrinos just before swallowed by the black hole.

After the black hole formation, the luminosities decrease slightly. 
However, a dense torus surrounding the black hole is formed in 
a short time scale. Then, the luminosity increases again, and 
becomes as large as the second and third peaks with the total 
luminosity $\sim 3 \times 10^{54}$ erg/s. 
Approximate generation rate of thermal energy at the shock on the surface of the torus 
due to infalling material is 
\beqn
\frac{GM_{\rm BH}\dot{M}}{r} &\sim& 0.1\dot{M}c^2 \nonumber \\
&\sim& 4 \times 10^{54} 
\left(\frac{\dot{M}}{20M_{\odot}~{s^{-1}}} \right)
\,{\rm erg/s}.
\eeqn 
Thus, the neutrinos are emitted by converting infall kinetic energy 
of the material to the thermal energy.

Convective motions are also observed in the rapidly model as in the moderately 
rotating model. A large-scale circulation is formed associated with
the formation of the thick, (mainly) centrifugally supported torus 
(see the bottom-middle panel in Figure \ref{fig_con-rho-r06}).
However, successive large-scale circulations, appeared in the moderately 
rotating model, do not occur in the rapidly rotating model,
although small-scale convective activities are driven
(see the bottom-right panel in Figure \ref{fig_con-rho-r06}).
This is due to the stabilizing effect of the epicyclic frequency
(see Eq. (\ref{Eq_fq})).
Figure \ref{fig_con-BV-r06} plots the Brunt-V\"ais\"al\"a frequency 
(see Eq. (\ref{Eq_BV})) and the Solberg-Hoiland frequency defined by Eq. (\ref{Eq_fq}).
As shown in this figure, there exist regions with negative gradients of 
entropy per baryon and lepton fraction ($N^{\ 2}_{\rm BV} < 0$) inside the 
thick torus (see the left panel in Figure \ref{fig_con-BV-r06}).
However, most of the low-frequency modes are suppressed by the
stabilizing epicyclic mode and only the higher-frequency modes are present.
Consequently, large-scale circulation modes are suppressed and
only small-scale convective modes appear.

Due to the absence of large-scale convective modes, effects of the convection 
on neutrino luminosities are likely to be minor.
Indeed, no violent time-variability is observed after the thick torus 
formation. The small bumps in luminosities at $t\approx 1500$--1510 ms are
associated with the large-scale circulation (see the bottom-middle panel in
Figure \ref{fig_con-rho-r06}).

The total mass of the torus is $\sim 7$\% of the black hole mass and gradually
increases (see Figure \ref{fig_BH-r06}).
The self-gravity of the torus may play a role in a later phase; 
the torus may be unstable against non-axisymmetric perturbation and 
this may affect evolution of
the torus because angular momentum transport and redistribution inside the 
torus are enhanced.
To strictly clarify the evolution of such massive torus, a three-dimensional
numerical simulation may be needed. This is one of the issues left for the
future.

Finally, we remark possible effects of viscosity in the rapidly rotating
model. Assuming that the torus can be described by the standard disk model, 
the mass accretion rate associated with a hypothetical viscous stress is 
estimated as $\dot{M}_{\rm vis} \sim 3$--$5M_{\odot}$ s$^{-1}$ for 
$\alpha_{\rm vis}=0.1$ (cf. Eq. (\ref{Eq_mdot_vis})).
Because the mass infalling rate onto the torus is
$\dot{M}_{\rm disk} \approx 8 M_{\odot}$ s$^{-1}$ at the late phase 
(see the bottom panel in Figure \ref{fig_BH-r06}), 
the viscosity is not expected to play a crucial role for the evolution of 
the torus at an early phase simulated in this paper.

However, in a later phase, when the mass infalling rate onto the torus becomes
smaller, the viscosity is expected to play an important role.
Then, an ADAF-type (accretion dominated accretion flow) accretion flow
may be the outcome in the presence of a large viscosity.
A high-velocity outflow may be accompanied because the accretion rate is likely to be
very high \citep[e.g.,][]{Narayan01}.
The high black hole spin may also play an important role for driving a high-velocity
outflow because the heating rate is enhanced near the ISCO and the mass accretion is
suppressed due to the small black hole radius.

\section{Discussions}\label{Sec_Discussion}

\subsection{Effect of the black hole spin on disk property and neutrino emissivity}
\label{Sec_spin}

The black hole formed after the core collapse
is in general not a Schwarzschild black hole but a rotating black hole 
($q_{\rm BH} \gtrsim 0.5$ for our models). 
In addition, a high spin state with $q_{\rm BH} \gtrsim 0.8$ is easily achieved
during the evolution of the black hole.
Thus, it is necessary to take into account the effects associated with such a 
high black hole spin to build plausible models in the collapsar scenario.

The spin of a black hole is known to play a crucial role on the evolution of 
the accretion disk \citep{Chen07}. 
The inner edge of the disk (or torus) around a rapidly rotating black hole comes closer 
to the black hole than that around a Schwarzschild black hole, and consequently, 
the temperature and density of the disk reach higher values. 
These significantly enhance neutrino luminosities. 
In addition, due to the higher density and temperature, the disk becomes 
more opaque to neutrinos, and neutrinos are often trapped in the inner regions 
of the disk. This leads to formation of regions with negative entropy gradient, 
and convection is induced. As a result of convection, neutrino luminosity curves
may become highly variable.

Here, it should be noted that the trapping of neutrinos and occurrence of 
convective motions are not likely to be special consequences of the high mass 
accretion rate ($\dot{M} \sim 10 M_{\odot}$ s$^{-1}$) achieved in our models.
According to results of a general relativistic study by \citet{Chen07}, 
the neutrino trapping occurs even with a moderate mass accretion rate of 
$\dot{M} \sim M_{\odot}$ s$^{-1}$ for accretion flows around a rapidly 
rotating Kerr black hole.
For accretion flows around a Schwarzschild black hole, by contrast, 
the neutrino trapping does not occur  even with a high mass accretion rate of 
$\dot{M} \sim 10 M_{\odot}$ s$^{-1}$ \citep{Chen07}.
This illustrates that the black hole spin plays a crucial role 
on the properties of accretion flows around a black hole.
They also find that the neutrino trapping occurs in the vicinity of
the black hole ($r \lesssim 20 GM_{\rm BH}/c^{2}$), as in our case.
This indicates the importance of resolving the regions in the vicinity of the
black hole because the seed of convection is formed there.
(We note that the enhancement of neutrino luminosities due to the convection 
was not found in previous pseudo-Newtonian studies because a rather wide region
near the black hole was excised in these studies.)

\subsection{Comparison with CDAFs}

\begin{figure}
  \begin{center}
    \epsscale{1.0}
    \plotone{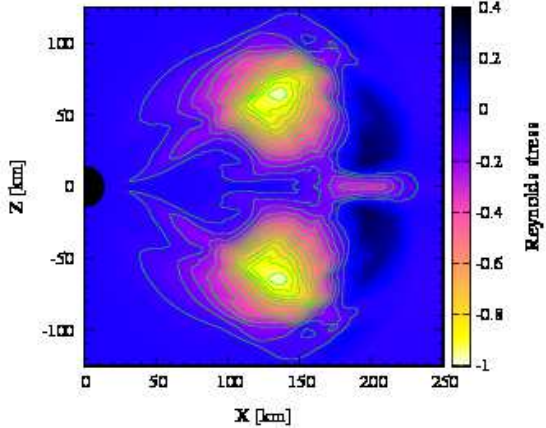}
  \end{center}
  \caption{The $r \varphi$-component of the Reynolds stress tensor ${\bm t}_{r\varphi}$ 
    normalized so that the maximum amplitude of negative sign is unity.
    \label{fig_Reynolds}}
\end{figure}

Presence of convective accretion flow, named as 
convection-dominated accretion flow (CDAF), was first 
predicted by \citet{Narayan94} in their studies of a 
self-similar solution of advection-dominated accretion 
flows (ADAFs). Later, CDAFs were found in numerical studies of ADAFs 
around a black hole \citep{Stone99,Igumenshchev00}.
They found as a remarkable property of CDAF that the convection
transports the angular momentum {\it inward} rather than outward.

To see whether this is the case in the present simulation, 
we calculate the $r\varphi$-component of the Reynolds stress tensor, 
${\bm t}_{r\varphi} = \langle \delta v_{r} \delta v_{\varphi} \rangle$,
where $\delta v_{i} = v_{i} - \langle v_{i} \rangle$ is the
velocity fluctuation and $\langle \ \ \rangle$ denotes time-averaging \citep{Igumenshchev00}. 
Note that negative (positive) sign of ${\bm t}_{r\varphi}$ corresponds 
to the inward (outward) transfer of the angular momentum.
Figure \ref{fig_Reynolds} plots contour of ${\bm t}_{r\varphi}$
in the $x$-$z$ plane. This figure clearly shows that there are regions with 
negative values of ${\bm t}_{r\varphi}$ near the outer surface of the torus. 
Convection in these regions transports the angular momentum inward, 
generating flows with higher angular momentum in an inner region.
Such flows will then move outward forming circulations.

While the CDAF-like accretion flows are formed in the outer part of
the torus, flows in the inner region are similar to those of 
neutrino-dominated accretion flows (NDAFs) \citep{Popham99}.
Furthermore, the torus is accompanied by the quasi-radial flows which 
consist of the material with low angular momentum and the outer geometrically thin accretion flows
near the equatorial plane.
\citet{Narayan01} found that transition between CDAF and
NDAF are determined by a characteristic radius $r_{\rm out}$:
Flows injected from $r\gtrsim r_{\rm out}$ form CDAFs,
and those injected from $r\lesssim r_{\rm out}$ form NDAFs.
In terms of the specific angular momentum, transition between CDAF and
NDAF may be determined by a characteristic specific angular momentum 
$j_{\rm out}$: The material with $j\gtrsim j_{\rm out}$ form CDAFs and those
with $j\lesssim j_{\rm out}$ form NDAFs.
As found in the present simulation, the accretion flows in 
the moderately rotating collapsar model
will be characterized by the inner NDAF-like and outer CDAF-like parts.

\subsection{Application to Gamma-ray bursts}\label{Sec_GRB}

We now turn to application of our results to LGRBs. 
We consider, as two possible ways of the energy deposition 
process, the neutrino pair annihilation and the Blandford-Znajek
process \citep{Blandford77}.
Because both of which processes are not included in our numerical 
simulation, we give an order estimate of the energy deposition rates 
for the purpose of clarifying the potential of driving relativistic
jets in our models.

The annihilation rate of neutrinos and anti-neutrinos into electron-positron pairs
has been calculated as a mechanism to power GRBs by several groups 
\citep{Ruffert97,Popham99,Asano00,Asano01,Salmonson01,Setiawan04,
Setiawan06,Birkl07,Harikae10a,Harikae10b,Zalamea11}.
The energy from the neutrino pair annihilation should be deposited
in a baryon-poor region in order to generate highly relativistic outflows.
The funnel region near the rotational axis above the torus is
a promising place for this purpose.

Here, we an order estimate of the total energy deposition rate by 
the neutrino pair annihilation ($\dot{E}_{\nu\bar{\nu}}$).
The deposition rate is proportional to $\dot{M}^{9/4}M_{\rm BH}^{-3/2}$ \citep{Beloborodov08}.
In this estimation, the neutrino luminosity is assumed to be originated from a
viscous heating. In our present calculation, the neutrino luminosity is determined by 
mass accretion rate of the infalling material which experiences the shock heating
and increases thermal energy of the disk. 
However, the dependence of the pair-annihilation rate on the mass infall rate 
$\dot{M}$ is essentially the same for thick torus phase.
Due to this strong dependence on the mass accretion rate, the energy deposition
by the neutrino pair-annihilation will be important only for an early phase of
the LGRB formation.

In the geometrically thin disk, the efficiency of the neutrino pair annihilation 
for a rapidly rotating black hole may be written, according to a recent 
general relativistic study by \citet{Zalamea11}, as
\beq
({\rm eff})_{\nu\bar{\nu}} \equiv \frac{\dot{E}_{\nu\bar{\nu}}}{L_{\nu,\,{\rm tot}}}
 \sim 0.01 \left( \frac{\dot{M}}{M_{\odot}\, {\rm s}^{-1}} \right)^{5/4}
\left( \frac{M_{\rm BH}}{10\,M_{\odot}}\right)^{-3/2},
\eeq
where $L_{\nu,\,{\rm tot}}$ is the total neutrino luminosity.
In the present simulation, the expected energy deposition rate by neutrino 
pair annihilation is quite high as $\dot{E}_{\nu\bar{\nu}}\sim 10^{53}$ erg/s for
$M_{\rm BH} \sim 10M_{\odot}$, $\dot{M} \sim 10M_{\odot}$ s$^{-1}$, and 
$L_{\nu,\, {\rm tot}} \sim 10^{54}$ erg/s in an early phase of disk evolution
for $\sim 1$ s.

The efficiency of the neutrino pair annihilation depends strongly on the geometry 
of the disk. In particular, (eff)$_{\nu\bar{\nu}}$ is proportional to 
$V^{-1}_{\rm ann}$, where $V_{\rm ann}$ is characteristic volume above the disk
\citep{Mochkovitch93,Liu10,Zalamea11}.
\citet{Liu10} calculated the vertical structure of 
geometrically-thick accretion flows in the pseudo-Newtonian gravity 
and estimated the energy deposition rate. 
They found that the efficiency could be enhanced by an order of magnitude.
In this case, a very large energy deposition rate by neutrino pair annihilation of 
$\dot{E}_{\nu\bar{\nu}}\sim 10^{54}$ erg/s may be expected. 

The outgoing Poynting power at the horizon in the Blandford-Znajek process 
is given by \citep{Blandford77,Thorne86}
\beq
\dot{E}_{\rm BZ}\approx  \frac{c}{32}\,q_{\rm BH}^{2}\,(B_{H}^{\perp})^{2}\,
R_{H}^{2}\, \frac{\Omega_{B}(\Omega_{H}-\Omega_{B})}{\Omega_{H}^{2}},
\label{Eq_BZ_pow}
\eeq
where $B_{H}^{\perp}$ is magnitude of magnetic fields normal to the horizon,
$R_{H}\sim GM_{\rm BH}/c^{2}$ is the radius of the horizon, 
and $\Omega_{H}$ and $\Omega_{B}$ are the angular velocities of the horizon 
and the magnetic field lines. 

\citet{McKinney05} suggested an approximate fitting formula for 
the estimation of the field strength based on results of general relativistic 
magnetohydrodynamical simulations. According to his formula, the outgoing Poynting 
power in the Blandford-Znajek process is given by
\beq
\dot{E}_{\rm BZ}
\sim 10^{52} \,f_{\Omega_{H}} \, q_{\rm BH}^{2}
\left( \frac{\dot{M}}{M_{\odot}\,{\rm s}^{-1}} \right) \,
{\rm erg/s} ,
\eeq
where $f_{\Omega_{H}}$ is a parameter which depends strongly on the angular 
velocity and the most optimistic condition $\Omega_{B}=\Omega_{H}/2$ is assumed.
According to the result of \citet{McKinney05}, 
$\gtrsim 10$\% of the total outgoing power may be used to produce the LGRB jet.
Thus, the outgoing jet power will be 
$\dot{E}_{\rm BZ,jet} \sim 10^{51} 
f_{\Omega_{H}}q^{2}_{\rm BH}(\dot{M}/(M_{\odot}\,{\rm s}^{-1}))$ erg/s for our models.

The Blandford-Znajek power will eventually become much larger than the deposition 
rate by the neutrino pair-annihilation because the power depends more weakly on the
mass accretion rate. Even in a late phase with $\dot{M}\sim 0.1M_{\odot}$ s$^{-1}$,
a jet power of $\dot{E}_{\rm BZ,jet} \sim 10^{51}$ erg/s may be achieved if 
the black hole is sufficiently rapidly rotating 
($q_{\rm BH} \gtrsim 0.9$ for which $f_{\Omega_{H}}\gtrsim 10$), 
accumulating the angular momentum of infalling material.
Note also that magnetic fields may be amplified in the torus due to the 
magneto-rotational instability and/or convection \citep{Balbus91,Balbus98}.

\subsection{Gravitational waves from anisotropic neutrino emission}

A cosmological population of core-collapse supernovae is one of the most important 
sources of gravitational wave backgrounds \citep{Buonanno05}.
Gravitational waves (GWs) associated with anisotropic neutrino emission
are particularly important because they generate a burst of GWs accompanying 
with the memory effect, the so-called burst with memory \citep{Braginskii87}. 
GW memory due to anisotropic neutrino emission could contaminate, 
at low frequencies around 0.1Hz, the inflationary GW 
\citep{Buonanno05,Hiramatsu05,Suwa07a}, 
which is one of the targets of future space GW detectors such as DECIGO 
\citep{Seto01} and BBO \citep{Ungarelli05}.
Here, we give an order estimate of the amplitude of GWs associated with anisotropic neutrino emission.

The amplitude of GWs due to anisotropic neutrino emission is given by 
\citep{Muller97,Kotake07,Suwa09}.
Taking characteristic values of total neutrino luminosity of $\sim 10^{54}$ erg/s from our 
simulation results and assuming a duration of neutrino emission of 
$\Delta t_{\nu} \sim 1$ s (cf. the moderately rotating model), the amplitude may be estimated as
\beq
h_{\nu} \sim 2 \times 10^{-24} 
\left( \frac{10 \, {\rm Gpc}}{D} \right)
\left( \frac{L_{\nu}}{10^{54}\,{\rm erg/s}} \right)
\left( \frac{\Delta t_{\nu}}{1\,{\rm s}} \right),
\eeq
where $D$ is the distance to the source.
This value is as large as that calculated by \citet{Suwa07a} for  
the collapse of $300 M_{\odot}$ PopIII stellar core collapse. 
Note that the initial core mass in \citet{Suwa07a} is about three times larger than ours.
The peak neutrino luminosities achieved in their results are by a factor of $\sim 10$ larger than
those in our results, while the duration in their results is by a factor of $\sim 10$ shorter than 
that in the moderately rotating model, because they failed to find convective activities 
in the accretion torus.

If long-term neutrino emission as found in the present simulations is universal 
for the Pop III stellar collapse, 
the GW memory due to anisotropic neutrino emission could significantly contaminate the 
inflationary GW.

\section{Summary}\label{Sec_Summary}

In this paper, we performed axisymmetric simulations of very
massive stellar core collapsing to a system composed of a rotating black hole and 
surrounding disk in full general relativity. 
We took into account a nuclear-theory-based finite-temperature EOS (Shen's EOS), weak interaction
processes such as electron capture and pair-neutrino processes, and neutrino cooling, which 
is handled by a general relativistic leakage scheme \citep{Sekiguchi10a,Sekiguchi10b}.

Progenitor models of LGRBs suggested in the literatures \citep[e.g.,][]{Fryer07}) raise a
possibility that they may have an entropy higher than that of ordinary supernova cores.
In this work, we employed a core with a high entropy of 
$s/k_B = 8$ as the initial condition. Because the distribution of angular
momentum in very massive stars is highly uncertain, we employed four models
(spherical, slowly rotating, moderately rotating, and rapidly rotating models) by 
superimposing a profile of rotational angular velocity in a parametric manner. 
The initial models adopted in this paper are not rapidly rotating in the sense that 
the rotation velocity imposed is much smaller than that required to retain the ISCO
around a Schwarzschild black hole and that considered in previous studies 
(e.g., \citet{MacFadyen99}, see also \citet{Lopez-Camara09} and references therein). 

As in the collapse of ordinary supernova cores, gravitational collapse sets in due to 
photo-dissociation of heavy nuclei and electron capture. However, the collapse dynamics and
properties of neutrino emission are different from those of ordinary supernova cores. The 
characteristics of the collapse of high-entropy cores are summarized as follows: 

\begin{enumerate}
\item The gravitational contraction is decelerated by the thermal gas-pressure of free nucleons
at a subnuclear density and the core experiences a weak bounce (the gas-pressure-dominated
bounce). This is a result of the high entropy. 
We reconfirmed this previous discovery \citep{Nakazato07,Suwa07b}
and clarified the physical origin in detail: We clarified that the weak bounce is universal for the
collapse of the core with $s/k_{B} \approx 5$--16.
\item Because the gas-pressure-dominated bounce is too weak to halt the infalling material, 
a black hole is formed soon after the bounce (within $\sim 30$ ms). The mass of the black hole 
at the moment of its formation ($\sim 5.8$--$7M_{\odot}$) is much larger than the maximum mass 
of a cold neutron star ($\approx 2.2M_{\odot}$for the Shen's EOS). 
This is also due to the high entropy (high thermal pressure). Just before the
black hole formation, the pair-neutrino production processes are enhanced because the
temperature increases due to the adiabatic compression (due to neutrino trapping).
As a result, approximately the same amount of electron neutrinos and anti-neutrinos are emitted.
The mass accretion rate into the black hole just after the black hole formation and 
the total neutrino luminosity just before the black hole formation
are $\sim 40 M_{\odot}$ s$^{-1}$ and $\sim 4 \times 10^{54}$ erg/s depending weakly on the degree of 
rotation. Thus the maximum efficiency for the neutrino emission is 
$L_{\nu}/(\dot{M}c^{2}) \sim 6$\%.
\item In the moderately rotating model, a geometrically thin accretion disk is first 
formed around the black hole and shocks are formed on its surface, generated by the 
infalling material. As the thermal energy is stored in the disk, it expands eventually 
to be a geometrically thick accretion torus. After the thick torus formation, 
convective activities, which are similar to those in CDAFs \citep{Narayan01}, set in 
because a region with negative entropy gradient emerges in the inner part of the torus,
due to occurrence of the neutrino trapping. The neutrino luminosities are
$L_{\nu_e} + L_{\bar \nu_e} \sim 10^{54}$ erg/s, and show violent time-variability. 
Here we emphasize that the source of thermal-energy generation, which is eventually
dissipated by neutrinos, is the shock heating of infalling materials.
The high spin of a black hole is likely to play a crucial role on the evolution of the accretion disk, 
convective activities, and the enhancement of neutrino luminosities. 
\item The evolution process of the accretion disk and neutrino emissivity depend strongly on the
degree of initial rotation. In the slowly rotating model, the disk remains geometrically
thin for a long time, and hence, the neutrino emissivity also remains relatively small 
($L \sim 10^{53}$ erg/s) for more than 100 ms. In the rapidly rotating model, by contrast, 
a geometrically thick torus is immediately formed after the black hole formation, and luminosities of 
neutrinos emitted from the torus are as high as $10^{54}$ erg/s even at its formation.
However, the convection is suppressed by the stabilizing epicyclic mode due to the rapid
rotation and no violent time-variability is observed in the neutrino luminosities.
\item Irrespective of the degree of rotation, long-lived disk or torus surrounding the black hole is a
primary emitter of neutrinos because of its high luminosity and long lifetime $\agt 1$ s.
This implies that anisotropic emission of neutrinos comes mainly from the accretion disk (torus)
surrounding a black hole, not from the dense matter collapsing to a black hole. For a correct
estimation of gravitational-wave background by anisotropic neutrino emission, it may be
necessary to understand the physical condition of the accretion disk or torus (see below).
\end{enumerate}

Finally, we comment on major limitations of the present study. 
First, we adopt initial conditions which are not based on latest theoretical 
models of stellar evolution.
We are going to perform simulations adopting more realistic initial models soon.
Second, the present simulations are performed on the assumption of axial symmetry.
The accretion disk formed in the present simulations may become unstable against
non-axisymmetric instabilities \citep[e.g.,][]{Korobkin11,Taylor11,Kiuchi11}. 
Competition between non-axisymmetric instabilities and convective instabilities 
should be explored.
Third, we do not take account of the neutrino heating.
A simple approximated procedure of including effects of neutrino heating
is adopted by \citet{O'Connor11} in which stellar core collapse to a black hole is 
studied by a spherically symmetric fully general relativistic simulation. 
We also plan to study effects of neutrino heating using a recently developed 
formulation \citep{SKSS11}.
Fourth, we do not consider effects of magnetic fields which will play a role during
the collapse \citep[e.g.,][]{Barkov08,Komissarov09} and subsequent evolution of
the disk \citep[e.g.,][]{Penna10,Barkov11} if progenitor cores have large magnetic fields.
We plan to perform simulations taking account of magnetic fields using a
general relativistic magnetohydrodynamic code we have developed \citep{Shibata05}.

\acknowledgments
We thank to K. Nakazato and K. Sumiyoshi for providing us extended hadronic EOS,
and to K. Ioka and Y. Suwa for valuable discussions and comments.
YS thanks to K. Ohsuga and M. Machida for valuable discussions.
He also thanks to T. Shiromizu and T. Fukushige for their grateful aids. 
Numerical computations were performed on the NEC SX-9 at the data analysis 
center of NAOJ and on the NEC SX-8 at YITP in Kyoto University. 
This work is supported by
the Grant-in-Aid for Scientific Research (21018008, 21105511, 21340051), 
and by the Grant-in-Aid for Scientific Research on Innovative Area (20105004) 
of Japanese MEXT.

\end{document}